\documentclass[tightenlines,eqsecnum,floats,aps,amsmath,amssymb,nofootinbib,prd,showpacs,notitlepage]{revtex4-1}
\pdfoutput=1 
\usepackage{amsmath,amssymb,amsfonts}
\usepackage{graphicx}
\usepackage{enumerate} 
\usepackage{colordvi} 
\usepackage{bm}
\usepackage{epsfig}
\usepackage{float}
\usepackage{verbatim}
\usepackage{subfig}
\usepackage{xcolor}

\newcommand{\bfx}{\mbox{\boldmath$x$}}
\newcommand{\bfk}{\mbox{\boldmath$k$}}
\newcommand{\bfp}{\mbox{\boldmath$p$}}
\newcommand{\bfq}{\mbox{\boldmath$q$}}

\begin{document}
\title{The one-loop matter bispectrum as a probe of gravity and dark energy}

\vfill
\author{Benjamin Bose$^{1}$, Atsushi Taruya$^{2,3}$}
\bigskip

\affiliation{$^1$ Yukawa Institute for Theoretical Physics, Kyoto University, Kyoto 606-8502, Japan}
\affiliation{$^2$ Center for Gravitational Physics, Yukawa Institute for Theoretical Physics, Kyoto University, Kyoto 606-8502, Japan}
\affiliation{$^3$ Kavli Institute for the Physics and Mathematics of the Universe (WPI), The University of Tokyo Institutes for Advanced Study, The University of Tokyo, 5-1-5 Kashiwanoha, Kashiwa, Chiba 277-8583, Japan}

%
\bigskip
\vfill
\date{today}
\begin{abstract}
Gravity-induced non-Gaussianity in the large-scale structure of the Universe, characterised by higher-order statistics such as the bispectrum (three-point cumulant), is expected to contain rich cosmological information. A measurement of the bispectrum will not only improve the cosmological constraints, but also give us the possibility to probe gravity on cosmological scales. In this paper, we present a framework to numerically calculate the one-loop matter bispectrum based on standard perturbation theory (SPT). This approach allows general modifications to the standard $\Lambda$CDM model to be easily implemented. We demonstrate the performance of the bispectrum calculation in three representative cases, namely the Vainshtein-screened Dvali-Gabadadze-Porrati (DGP) model, the chameleon-screened Hu-Sawicki $f(R)$ model and the phenomenological dark scattering (DS) momentum-exchange model. The predicted bispectra are then compared with measured results from a set of cosmological $N$-body simulations, and the impact of possible systematics arising from simplified or approximate treatments in the perturbative calculation is studied in detail. We find that the one-loop bispectrum calculation offers significantly more information on general screening and momentum exchange effects than the leading-order bispectrum calculation. Further, the accuracy of the one-loop prediction is shown to be comparable to non-linear fitting formulas over a wide range of wavenumbers ($k\lesssim0.3\,h$$\mbox{Mpc}^{-1}$) even at lower redshifts, $z\lesssim 1$. 
\end{abstract}
\maketitle

\section{Introduction}
\label{sec:intro}


The concordance model of cosmology, i.e. general relativity (GR) with  constant dark energy ($\Lambda$) and cold dark matter (CDM) components, is now widely accepted as the most successful cosmological model. Indeed, with only $6$ parameters the model consistently describes both cosmic expansion and structure formation and accommodates not only the high-precision data set of the cosmic microwave background (CMB) \cite{Planck:2015xua} but also various measurements made of the late-time universe such as cluster counts \cite{Hamana:2015bwa}, baryon acoustic oscillations (BAO) \cite{Anderson:2013zyy} and supernovae data  \cite{Lampeitl:2009jq}. However despite its great success our understanding of the Universe is still limited. The concordance model implies that the Universe's geometry is close to flat and that it is filled with the hypothetical CDM, together with a small fraction of baryons. Moreover, the $\Lambda$CDM model assumes an unknown energy component called dark energy which is the underlying cause of the observed late time acceleration of the Universe \cite{Riess:1998cb,Perlmutter:1998np}. The dark energy may be explained by the non-zero cosmological constant $\Lambda$, but its smallness leads to the biggest fine-tuning problem in fundamental physics \cite{Weinberg:1988cp,Martin:2012bt}. Further, several tensions in cosmological parameters between local/low-$z$ measurements and CMB data have been recently advocated, specifically with respect to the present-day Hubble constant $H_0$ \cite{Efstathiou:2013via,Zhang:2017aqn,Riess:2009pu} and amplitude of density fluctuations $\sigma_8$ \cite{Abbott:2017wau,Beutler:2016arn} (see \cite{Lin:2017bhs} for a review). These problems may suggest that the underlying assumption of GR in the $\Lambda$CDM model is wrong and gravity is modified at cosmological scales (see \cite{Koyama:2015vza} for a review). Also, as an alternative scenario, the cosmological constant may be replaced with a dynamical dark energy with potential interactions with the dark matter sector (see \cite{Yoo:2012ug} for a review). 

Modified gravity (MG) has been often invoked in order to explain the accelerated expansion, introducing extra degrees of freedom. Most of MG models involve a scalar field which generally results in additional forces and hence modifies the gravitational force predicted by GR. A crucial point is that in order for such models to be viable, so-called screening mechanisms, by which the theory recovers GR at small scales, need to be self-consistently implemented. Hu-Sawicki $f(R)$ gravity \cite{Hu:2007nk} and Dvali-Gabadadze-Porrati (DGP) \cite{Dvali:2000hr} models are prototypical examples having such mechanisms. Recently, larger classes of healthy models has been uncovered, referred to as the Horndeski class \cite{Horndeski:1974wa}, beyond Horndeski \cite{Gleyzes:2014dya} and extended scalar-tensor theories \cite{Crisostomi:2016czh}. 

On the other hand, if we choose to accept the idea of dark energy instead of MG, there is no reason to stop us from considering departures from a pure cosmological constant. Such modifications are again described by introducing free parameters, and one simple example is the equation-of-state parameter which changes the cosmic expansion at late times. One may also consider the interaction within the dark sector, and introduce energy or momentum exchange between dark matter and dark energy in a parametric form \cite{Simpson:2010vh,Lesgourgues:2015wza,Pourtsidou:2016ico,Baldi:2016zom,Buen-Abad:2017gxg}. These theories must retain all the observational successes of the $\Lambda$CDM model. A particularly interesting alternative to the cosmological constant may be the case of momentum exchange between dark energy and dark matter which has a general formulation at the Lagrangian level \cite{Pourtsidou:2013nha}. This has been shown to explain the CMB as well as to weaken the tensions in the $\sigma_8$ parameter \cite{Pourtsidou:2016ico}. 

There are thus various possible alternatives to $\Lambda$CDM which should be tested against future precision observations, especially at cosmological scales. In this respect, galaxy redshift surveys and weak lensing experiments offer nearly ideal testing grounds, and with future stage-IV class surveys such as EUCLID \footnote{\url{www.euclid-ec.org}} \cite{Laureijs:2011gra}, WFIRST \footnote{\url{https://wfirst.gsfc.nasa.gov/}} \cite{Spergel:2013tha}, DESI\footnote{\url{http://desi.lbl.gov/}}  \cite{Aghamousa:2016zmz} and LSST\footnote{\url{https://www.lsst.org/}} \cite{Chang:2013xja}, we will be able to falsify or detect any deviation from $\Lambda$CDM at an unprecedented level. To make the best use of the statistical precision data, theoretical descriptions of the large-scale structure must be improved, accounting for any observational systematics including non-linear gravitational evolution. This is indeed essential to extract vital and non-degenerate information about the gravitational potential \cite{Baker:2014zba} and is the subject of active research \cite{Baumann:2010tm,Perko:2016puo,Lewandowski:2016yce,Schneider:2010gv,Heitmann:2006hr,Hashimoto:2017klo,Schmittfull:2016yqx}. 

If we are to move toward unbiased and improved tests of gravity and dark energy, future high-precision data not only requires us to carefully quantify the accuracy of theoretical templates \cite{Bose:2018orj,Bose:2017myh,Barreira:2016ovx,Taruya:2010mx,Taruya:2016jdt,Lewandowski:2017kes}, but also prompts us to use higher-order statistics such as the bispectrum or the three-point correlation function as informative cosmological signals, which will be measured at high-statistical significance. On top of the traditional method using two-point statistics, adding a bispectrum measurement is expected to improve the constraints on gravity and cosmology \cite{Child:2018klv,Byun:2017fkz,Song:2015gca}. Also in \cite{An:2017kqu} the authors show that weak lensing tomography is very sensitive to energy exchange in the dark sector and that the bispectrum can provide tighter constraints over the conventional convergence power spectrum. Further, \cite{Namikawa:2018erh}  shows that the CMB lensing bispectrum can be used to get clean constraints on general MG theories. Note, however, that while there have been numerous works on modeling the bispectrum in alternative theories of gravity \cite{Hirano:2018uar,GilMarin:2011xq,Yamauchi:2017ibz,Bellini:2015wfa,An:2017kqu,Dinda:2018eyt}, most of the analytic works are restricted to a leading-order calculation only valid at very large scales. On the issue of moving to the non-linear small scales, numerical simulations are still a computationally expensive and impractical approach in the context of survey data analyses. 

In this paper, we try to fill the gap between the leading-order analytic calculation and fully non-linear simulations by employing the next-to-leading order perturbative calculation in alternatives to $\Lambda$CDM. To be precise, employing the numerical algorithm described in \cite{Taruya:2016jdt}, we extend the power spectrum code presented in \cite{Bose:2016qun} to compute the matter bispectrum at one-loop order. Based on the newly developed code, we demonstrate the one-loop predictions of the bispectrum in three representative models: Vainshtein screened DGP \cite{Dvali:2000hr} model, the Hu-Sawicki $f(R)$ chameleon screened model \cite{Hu:2007nk} and the dark scattering (DS) momentum exchange model \cite{Simpson:2010vh,Baldi:2016zom}. The present code can be easily extended to a wide class of alternative models, for example the Horndeski class of MG theories with a generalised potential \cite{Bose:2016qun} or general dark energy models. We also highlight the power of the bispectrum for distinguishing between alternatives and $\Lambda$CDM. In particular we investigate the signal of one-loop contributions from screening or interaction effects. Further, we will compare the one-loop computation with another promising  non-linear prescription for the matter bispectrum in order to identify optimal theoretical frameworks for next generation analyses pipelines. 

This paper is organised as follows:  Sec.II presents the generalised evolution equations for the density perturbations and the expressions for the one-loop statistics. We describe modifications coming from three representative  non-standard models, namely  DGP, $f(R)$ and the DS model. Further, we highlight the numerical treatment of the perturbations used in this work. In  Sec.III  we test the perturbative predictions against sets of numerical simulations. We also compare our numerical PT approach against common approximations and other non-linear prescriptions for the bispectrum. In  Sec IV we investigate the non-linear signal of MG's dependence on bispectrum shape and redshift. Finally,  Sec.V gives a summary of the results and discusses future work. 


\section{Theory}
\label{sec:theory}
In this paper, we are interested in constructing two statistical quantities relevant for large-scale structure observations, the power spectrum and bispectrum. We shall compute these quantities based on standard perturbation theory (SPT), starting with Gaussian initial conditions. We will work far inside the Hubble horizon and so can safely ignore relativistic corrections, but we consider large enough scales so that non-linear effects of gravity are mild (the \emph{Newtonian regime}). The background cosmic expansion is assumed to follow $\Lambda$CDM model, but the perturbations will be treated generally. Further, the metric perturbations are assumed to be varying slowly with time and so time derivatives will be ignored in our treatment (the \emph{quasi-static approximation}).

\subsection{Perturbative framework}

In what follows, based on \cite{Bernardeau:2001qr}, we describe our basic formalism to treat the evolution of matter fluctuations. We consider scalar perturbations around the Friedmann-Lema\^itre-Robertson-Walker metric, which are expressed in Newtonian gauge as
\begin{equation}
ds^2=-(1+2\Phi)dt^2+a(t)^2(1-2\Psi)\delta_{ij}dx^idx^j, 
\end{equation}
with the function $a$ being the scale factor of the Universe. The background cosmic expansion of this metric is described by the Friedmann equation:
\begin{equation}
 \left(\frac{\dot{a}}{a}\right)^2\equiv H^2(a)=
H_0^2 \left[\Omega_{m,0} a^{-3} + \Omega_{{\rm DE},0}\exp\left\{\int^a_13[1+w(\tilde{a})]\tilde{a}d\tilde{a}\right\}\,\right],
\end{equation}
where $H_0$ is the present-day value of the Hubble parameter, $\Omega_{{\rm DE},0}$ and $\Omega_{\rm m,0}$ are the present-day density parameters of dark energy and dark matter, respectively. The function $w(a)$ represents the equation-of-state parameter of dark energy. Setting $w=-1$, the above equation is reduced to the Friedmann equation in the $\Lambda$CDM model. 

We are interested in large scales where the matter fluctuations can be described by the collisionless Boltzmann equation under the so called single-stream approximation. This is especially true for the early stages of structure formation. Then, the evolution of CDM and baryon fluctuations can be regarded as an irrotational and pressureless single-fluid system. Although the single-stream approximation is eventually violated in the non-linear regime at small scales, we shall keep relying on this treatment in predicting observables at large scales in generalised cosmologies. Then, the relevant quantities for evolution of fluctuations to be solved are the density field ($\delta$) and velocity-divergence field ($\theta$), defined as follows
\begin{equation}
\delta({\bf x})=\frac{\rho_{\rm m}({\bf x})-\bar{\rho}}{\bar{\rho}},
\qquad
\theta({\bf x})=\frac{\nabla\cdot \mbox{\boldmath$v$}({\bfx})}{aH(a)}.
\end{equation}
The evolution equations for these quantities, under the quasi-static treatment of metric and scalar field perturbations, are given in Fourier space by 
(e.g. \cite{Koyama:2009me,Taruya:2016jdt,Bose:2016qun})  
\begin{eqnarray}
&&a \frac{\partial \delta(\bfk)}{\partial a}+\theta(\bfk) =-
\int\frac{d^3\bfk_1d^3\bfk_2}{(2\pi)^3}\delta_{\rm D}(\bfk-\bfk_{12})
\alpha(\bfk_1,\bfk_2)\,\theta(\bfk_1)\delta(\bfk_2),
\label{eq:Perturb1}\\
&& a \frac{\partial \theta(\bfk)}{\partial a}+
\left(2+ A(a) + \frac{a H'}{H}\right)\theta(\bfk)
-\left(\frac{k}{a\,H}\right)^2\,  \Phi(\bfk)= \nonumber \\ &&
-\frac{1}{2}\int\frac{d^3\bfk_1d^3\bfk_2}{(2\pi)^3}
\delta_{\rm D}(\bfk-\bfk_{12})
\beta(\bfk_1,\bfk_2)\,\theta(\bfk_1)\theta(\bfk_2),
\label{eq:Perturb2}
\end{eqnarray}
where a prime denotes a scale factor derivative and $\boldsymbol{k}_{1...n} = \boldsymbol{k}_1+\cdots+\boldsymbol{k}_n$. The functions $\alpha$ and $\beta$ are the mode-coupling kernels given by 
\begin{eqnarray}
\alpha(\bfk_1,\bfk_2)=1+\frac{\bfk_1\cdot\bfk_2}{|\bfk_1|^2},
\quad\quad
\beta(\bfk_1,\bfk_2)=
\frac{(\bfk_1\cdot\bfk_2)\left|\bfk_1+\bfk_2\right|^2}{|\bfk_1|^2|\bfk_2|^2}.
\label{alphabeta}
\end{eqnarray}
At the level of generality addressed in this paper, we have included a drag term $A(a)$ ($A(a)=0$ in $\Lambda$CDM) in Eq.(\ref{eq:Perturb2}) which we discuss in the next subsection.  Further, in the context of MG, the Newtonian potential $\Phi$  is governed by a modified Poisson equation. In Fourier space, this reads \cite{Koyama:2009me} 
\begin{equation}
-\left(\frac{k}{a H(a)}\right)^2\Phi (\bfk;a)=
\frac{3 \Omega_m(a)}{2} {\mu(k;a)} \,\delta(\bfk) + { S(\bfk;a)},
\label{eq:poisson1}
\end{equation}
where $\Omega_m(a) =\kappa \rho_m/3 H^2$ and $\kappa = 8\pi G_N$, where $G_N$ is Newton's gravitational constant. $\mu(k;a)$ is the linear modification to gravity and is unity in the case of GR. The non-linear source term $S(\bfk ; a )$ characterizes new mode couplings, including those responsible for screening effects. In GR $S(\bfk;a) = 0$ but in general, up to fourth order in the perturbations, it is given by
\begin{eqnarray}
S(\bfk;a)&=&
\int\frac{d^3\bfk_1d^3\bfk_2}{(2\pi)^3}\,
\delta_{\rm D}(\bfk-\bfk_{12}) \gamma_2(\bfk_1, \bfk_2;a)
\delta(\bfk_1)\,\delta(\bfk_2),
\nonumber\\
&& + 
\int\frac{d^3\bfk_1d^3\bfk_2d^3\bfk_3}{(2\pi)^6}
\delta_{\rm D}(\bfk-\bfk_{123})
\gamma_3(\bfk_1, \bfk_2, \bfk_3;a)
\delta(\bfk_1)\,\delta(\bfk_2)\,\delta(\bfk_3) \nonumber \\ 
&& + 
\int\frac{d^3\bfk_1d^3\bfk_2d^3\bfk_3 d^3 \bfk_4}{(2\pi)^9}
\delta_{\rm D}(\bfk-\bfk_{1234})
\gamma_4(\bfk_1, \bfk_2, \bfk_3 , \bfk_4;a)
\delta(\bfk_1)\,\delta(\bfk_2)\,\delta(\bfk_3)\delta(\bfk_4).   \nonumber 
\label{eq:Perturb3}
\end{eqnarray}
We present specific forms for $A(a)$, $\mu(k;a)$ and $\gamma_i$ in the next subsection.

Provided the basic equations for perturbations [i.e. Eqs.~(\ref{eq:Perturb1}) and Eq.(\ref{eq:Perturb2})], the approach of SPT is to expand $\delta$ and $\theta$, and to solve them order by order. Our focus is the matter fluctuations seeded by tiny density fluctuations at early times, $\delta_0$. In this case, the $n^{th}$ order solutions are expressed as
\begin{align} 
\delta_n(\boldsymbol{k};a) &= \frac{1}{(2\pi)^{3(n-1)}}\int d^3\boldsymbol{k}_1...d^3 \boldsymbol{k}_n \delta_D(\boldsymbol{k}-\boldsymbol{k}_{1...n}) F_n(\boldsymbol{k}_1,...,\boldsymbol{k}_n, a) \delta_0(\boldsymbol{k}_1)...\delta_0(\boldsymbol{k}_n) \label{nth1} \\ 
\theta_n(\boldsymbol{k};a) &= \frac{1}{(2\pi)^{3(n-1)}}\int d^3\boldsymbol{k}_1...d^3 \boldsymbol{k}_n \delta_D(\boldsymbol{k}-\boldsymbol{k}_{1...n}) G_n(\boldsymbol{k}_1,...,\boldsymbol{k}_n, a) \delta_0(\boldsymbol{k}_1)...\delta_0(\boldsymbol{k}_2) \label{nth2},
\end{align}
where $F_i(\bfk_1,\bfk_2...,\bfk_i;a)$ and $G_i(\bfk_1,\bfk_2...,\bfk_i;a)$ are the $i^{th}$ order SPT kernels. Recalling that the random field $\delta_0$ follows Gaussian statistics, the matter power spectrum and bispectrum at next-to-leading order, called one-loop, can be calculated using the kernels up to fourth order. Their expressions are given by 
\begin{align} 
P^{1-{\rm loop}}(k;a)  = & P^{11}(k;a) \nonumber \\ & + P^{22}(k;a) + P^{13}(k;a), \label{1loopps} \\
B^{1-{\rm loop}}(k_1,k_2,\theta;a) = &  B^{112}(k_1,k_2,\theta;a) \nonumber \\ & +B^{222}(k_1,k_2,\theta;a)  +  B^{321}(k_1,k_2,\theta;a) + B^{114}(k_1,k_2,\theta;a) \label{1loopbs},
\end{align} 
where $\theta = \cos^{-1}{(\hat{\bfk}_1\cdot \hat{\bfk}_2)}$ \footnote{$\theta$ should not be confused with the velocity perturbation, $\theta_i(\bfk)$, which always appears with its arguments and subscript.} and we use the usual definitions
\begin{align}
\langle \delta_{n}(\bfk) \delta_m(\bfk')\rangle &=
(2\pi)^3\delta_{\rm D}(\bfk+\bfk')\,P^{nm}(k), \label{eq:psconstraint0} \\
\langle \delta_{n}(\bfk_1) \delta_m(\bfk_2) \delta_o(\bfk_3) \rangle &=
(2\pi)^3\delta_{\rm D}(\bfk_1+\bfk_2 + \bfk_3 )\,B^{nmo}(\bfk_1,\bfk_2,\bfk_3),
\end{align}
where $\delta_n$ is the $n^{th}$ order perturbation and we must add all permutations on the LHS, for example $B_{111}^{114} \sim \langle \delta_4 \delta_1 \delta_1 + \delta_1 \delta_4 \delta_1 + \delta_1 \delta_1 \delta_4 \rangle$. We  can now present the following expressions written explicitly in terms of the integral kernels $F_i$ 
\begin{align}
P^{11}(k;a) &= F_1(k;a)^2 P_L(k), \label{tree1} \\
P^{22}(k;a) &= \int \frac{d^3p}{(2\pi)^3} F_2(\bfp,\bfk-\bfp;a)^2 P_L(p)P_L(|\bfk-\bfp|), \\ 
P^{13}(k;a) &= 2F_1(k;a)P_L(k) \int \frac{d^3p}{(2\pi)^3} F_3(\bfp,-\bfp,\bfk;a) P_L(p), \\ 
B^{112}(k_1,k_2,\theta;a) & = 2 \Big[ F_2(\bfk_1,\bfk_2;a)F_1(\bfk_1;a)F_1(\bfk_2;a) P_L(k_1) P_L(k_2) + 2 \mbox{perms} (\bfk_1 \leftrightarrow \bfk_2 \leftrightarrow \bfk_3) \Big], \label{tree2} \\ 
B^{222}(k_1,k_2,\theta;a) & = 8 \int \frac{d^3p}{(2\pi^3)} F_2(\bfp,\bfk_1-\bfp;a) F_2(-\bfp,\bfk_2+\bfp;a)  F_2(-\bfk_1+\bfp,-\bfk_2-\bfp;a) \nonumber \\ &\times   P_L(p) P_L(|\bf k_1-\bfp|)P_L(|\bfk_2+\bfp|), \label{222comp}\\ 
B^{321-I}(k_1,k_2,\theta;a) & = 6 \Big[ F_1(\bfk_1;a)P_L(k_1) \int \frac{d^3p}{(2\pi^3)} F_2(\bfp,\bfk_2-\bfp;a) F_3(-\bfk_1,-\bfp,-\bfk_2+\bfp;a) \nonumber \\ & \times P_L(p) P_L(|\bfk_2-\bfp|)   +  \mbox{5 perms} (\bfk_1 \leftrightarrow \bfk_2 \leftrightarrow \bfk_3) \Big], \\ 
B^{321-II}(k_1,k_2,\theta;a) & =6\Big[ F_1(\bfk_1;a) F_2(\bfk_1,\bfk_2;a) P_L(k_1) P_L(k_2) \int \frac{d^3p}{(2\pi^3)}  F_3(\bfk_2,\bfp,-\bfp;a) P_L(p) \nonumber \\ &  +  \mbox{5 perms} (\bfk_1 \leftrightarrow \bfk_2 \leftrightarrow \bfk_3) \Big],\\ 
B^{411}(k_1,k_2,\theta;a) & = 12 \Big[F_1(\bfk_1;a) F_1(\bfk_2;a) P_L(k_1) P_L(k_2) \int \frac{d^3p}{(2\pi^3)}  F_4(-\bfk_2, -\bfk_1,\bfp,-\bfp;a) P_L(p) \nonumber \\ &  +  \mbox{2 perms} (\bfk_1 \leftrightarrow \bfk_2 \leftrightarrow \bfk_3)\Big], \label{fourtho}
\end{align}
where $\bfk_3 = -\bfk_1 - \bfk_2$. 

\subsection{Specific model examples}  
\label{subsec:models}
 
As specific examples we consider three alternatives to $\Lambda$CDM, here giving explicit forms for the functions $A(a)$, $\mu(k;a)$ and $\gamma_i$ (for $i \in \{ 2,3,4 \}$) which appear in Eq.~(\ref{eq:Perturb2}) and Eq.~(\ref{eq:poisson1}); the normal branch of DGP (nDGP), Hu-Sawicki $f(R)$ gravity, and the phenomenological dark scattering model. Note that these are chosen just as representative examples and our numerical procedure is quite general and any scalar-tensor theory or non-standard dark sector model can be implemented in principle (see \cite{Bose:2016qun} for example). 

\subsubsection{nDGP gravity}
\label{subsubsec:nDGP}

The DGP model of gravity \cite{Dvali:2000hr} assumes we live on a 4-dimensional manifold embedded in a 5D spacetime called the bulk.  At the time, this theory gained a lot of attention for not requiring a cosmological constant to explain cosmic acceleration. It does this by having gravity `dilute' at large distances through the 5th dimension. The DGP action can be written as follows
\begin{equation}
S_{\rm DGP} = \frac{1}{32 \pi r_c} \int d^5x \sqrt{-g_5} R_5 + \int d^4 x \sqrt{-g}( \frac{R}{2 \kappa} + L_M),
\label{DGP}
\end{equation}
where $R_5$ and $g_5$ are the Ricci Scalar and metric in 5D, while $L_M$ is the matter Lagrangian confined to the 4D manifold. $r_c$ is the model's free parameter which represents the scale at which we cross from the 4D gravity to the 5D gravity regime. Applying this model to a FLRW cosmology we obtain the Friedman equation  
\begin{equation}
\epsilon \frac{H}{r_c} = H^2 - \frac{\kappa}{3} \rho_m,
\label{DGPfried}
\end{equation}
where $\epsilon = \pm 1$ .  The $+$ solution provided the attractive alternative to $\Lambda$ by offering a self-accelerating solution. This branch was found to be theoretically unviable, or `ghostly'.  On the other hand, the $-$ solution (nDGP) is theoretically healthy but requires a cosmological constant to achieve acceleration at late times. This model is interesting nevertheless because of its screening properties as well as accurate analytic solutions to the evolution equations (see Appendix B). The function $\mu(k;a)$ characterising the linear modifications to the clustering equations is given in nDGP by 
\begin{equation}
\mu(k;a) \equiv 1 + \frac{1}{3\beta}  \ , \qquad 
 \label{betadef}
\beta(a) \equiv 1+\frac{H}{H_0} \frac{1}{\sqrt{\Omega_{rc}}}   \left(1+\frac{aH'}{3H}\right) \ .
\end{equation}
Note $\beta(a)$ should not be confused with the mode coupling kernel $\beta(\bfk_1,\bfk_2)$ which can be distinguished by its scale dependency. Here we choose to parameterize the cross-over scale in terms of $\Omega_{rc} \equiv 1/(4r_c^2H_0^2)$. The higher order coupling kernels are given by \cite{Bose:2016qun}
\begin{equation}
\gamma_2(\bfk_1,\bfk_2;a) = -\frac{H_0^2}{24 H^2 \beta(a)^3 \Omega_{rc}} \left(\frac{\Omega_{m0}}{a^3}\right)^2 (1-\mu_{1,2}^2) ,
\end{equation}
\begin{equation}
\gamma_3(\bfk_1,\bfk_2,\bfk_3;a) = \frac{H_0^2}{144 H^2 \beta(a)^5 \Omega_{rc}^2} \left(\frac{\Omega_{m0}}{a^3}\right)^3 (1-\mu_{2,3}^2) (1-\mu_{1,23}^2),
\end{equation}
and the fourth order contribution is given by \cite{Taruya:2014faa}
\begin{align}
\gamma_4(\bfk_1,\bfk_2,\bfk_3,\bfk_4;a) &= -\frac{H_0^2}{3456 H^2 \beta(a)^7 \Omega_{rc}^3} \left(\frac{\Omega_{m0}}{a^3}\right)^4  
\nonumber \\ & \times \left[(1-\mu_{1,2}^2) (1-\mu_{3,4}^2) (1-\mu_{12,34}^2) + 4 (1-\mu_{234,1}^2)(1-\mu_{3,4}^2) (1-\mu_{34,2}^2) \right],
\end{align}
where $\mu_{i,j} = \hat{\bfk_i}\cdot \hat{\bfk_j}$ is the cosine of the angle between $\bfk_i$ and $\bfk_j$.

\subsubsection{Hu-Sawicki $f(R)$ gravity}
\label{subsubsec:fR_gravity}

$f(R)$ gravity is a class of models in which the Einstein-Hilbert action is generalised to include an arbitrary function of the scalar curvature. Among various examples for the functional form of $f(R)$,  the Hu-Sawicki model \cite{Hu:2007nk} is well-studied  \cite{Song:2015oza,Hammami:2015iwa,Lombriser:2013wta,Hellwing2013,Zhao:2013dza,Okada:2012mn,Li:2011pj,Lombriser:2010mp,Schmidt:2009am,Brax:2008hh,Song:2007da,Burrage:2017qrf},  and provides a simple form with which chameleon-type screening is realised. It is given by 

\begin{equation}
f(R)  = -m^2 \frac{c_1 (R/m^2)^n}{c_2(R/m^2)^n+1}.
\label{husawicki}
\end{equation}
In this paper, we specifically consider the $n=1$ case. That is, the above equation is reduced to 
\begin{equation}
f(R)  \propto \frac{R}{AR+1},
\end{equation} 
with $A$ being a constant with dimensions of length squared. In the regime we are interested in, that is the high curvature regime, $AR >> 1$ we can expand $f(R)$ as 
\begin{equation}
f(R) \simeq -2\kappa \rho_\Lambda - f_{R0} \frac{R_0^2}{R},
\end{equation}
where $\rho_\Lambda$ depends on $A$, $R_0$ is the background curvature today. We have defined $f_{R0} \equiv \bar{f}_R(R_0)$, the bar indicating it is evaluated on the background. $|\bar{f}_{R0}|$ is the free parameter of the theory. When $|\bar{f}_{R0}|\ll1$, the background cosmology becomes indistinguishable with  $\Lambda$CDM, and we have
\begin{equation}
R_0 = H_0^2(12-9\Omega_{m0}).
\end{equation}
Using the above relations and the $f(R)$ form of the Poisson equation (see \cite{Koyama:2009me,Taruya:2014faa} for example), we can compare with Eq.~(\ref{eq:poisson1}) to get following non-linear interaction terms  
 \begin{align}
\mu(k;a) = &1 + \left(\frac{k}{a}\right)^2\frac{1}{3\Pi(k;a)}, \\ 
\gamma_2(\bfk_1,\bfk_2;a)  = &- \frac{3}{16}\left(\frac{kH_0}{aH}\right)^2\left(\frac{\Omega_{m,0}}{a^3}\right)^2  \frac{\Xi(a)^5}{f_0^2 (3\Omega_{m,0}-4)^4}\frac{1}{\Pi(k;a)\Pi(k_1;a)\Pi(k_2;a)}, \label{frg2} \\ 
\gamma_3(\bfk_1,\bfk_2,\bfk_3;a)   = &  \frac{1}{32} \left(\frac{kH_0}{aH}\right)^2 \left(\frac{\Omega_{m,0}}{a^3}\right)^3 \frac{1}{\Pi(k;a)\Pi(k_1;a)\Pi(k_2;a)\Pi(k_3;a)}  \nonumber \\ 
 \times  & \left[-5\frac{\Xi(a)^7}{f_0^3(3\Omega_{m,0}-4)^6} + \frac{9}{2}\frac{1}{\Pi(k_{23};a) }\left( \frac{\Xi(a)^5}{ f_0^2 (3\Omega_{m,0}-4)^4} \right)^2\right],\label{frg3} \\ 
 \gamma_4(\bfk_1,\bfk_2,\bfk_3,\bfk_4;a)   = &- \frac{1}{256}\left(\frac{kH_0}{aH}\right)^2 \left(\frac{\Omega_{m,0}}{a^3}\right)^4 \frac{1}{\Pi(k;a)\Pi(k_1;a)\Pi(k_2;a)\Pi(k_3;a) \Pi(k_{4};a)}  \nonumber \\ 
 \times  & \Big[35 \frac{ \Xi(a)^9}{f_0^4 (4-3\Omega_{m,0})^8 } + \frac{27}{4}\frac{ \Xi(a)^{15}}{ f_0^6 (4-3\Omega_{m,0})^{12 } \Pi(k_{12};a) \Pi(k_{34};a)} \nonumber \\ & + 45 \frac{\Xi(a)^{12}}{ f_0^5 (4-3\Omega_{m,0})^{10} \Pi(k_{12};a)} +   54 \frac{ \Xi(a)^{15} }{  f_0^6 (4-3\Omega_{m,0})^{12}  \Pi(k_{123};a) \Pi(k_{12};a)} \nonumber \\ & + 30 \frac{\Xi(a)^{12} }{f_0^5 (4-3\Omega_{m,0})^{10} \Pi(k_{123};a)} \Big], \label{frg4} 
\end{align}
where
\begin{equation}
\Pi(k;a) = \left(\frac{k}{a}\right)^2+\frac{\Xi(a)^3}{2f_0(3\Omega_{m,0}-4)^2}, \qquad  \Xi(a) =  \frac{\Omega_{m,0}+4a^3(1-\Omega_{m,0})}{ a^3},
\end{equation}
and $f_0 = |\bar{f}_{R0}|/H_0^2$.

\subsubsection{Dark scattering interaction model} 
\label{subsubsec:Interacting_DE}

Among various proposed models of dark energy having interactions in the dark sector, we consider the dark scattering (DS) model of \cite{Simpson:2010vh,Baldi:2016zom}. This phenomenological model aims to describe an elastic scattering between dark matter and dark energy, giving rise to only momentum exchange in the dark sector. Since there is no other channel of interaction, only Eq.(\ref{eq:Perturb2}) is modified, coming in the form of 
\begin{equation}
A(a) \equiv [1+w(a)]\frac{H_0^2}{H}\frac{3\xi}{\kappa}\Omega_{DE,0}\exp \Big[{\int^a_1 \frac{3[1+w(a)]}{\tilde{a}} d\tilde{a}} \Big] \, ,
\label{interactionterm}
\end{equation}
where $\xi$ quantifies the magnitude of the drag force arising from scattering and will be quoted in units of [bn $\mbox{GeV}^{-1}$]. We can now see that the term $A$ can act to oppose or enhance the evolution of velocity perturbations depending on whether $w$ is above or below the cosmological constant value $w=-1$. Further, $\mu(k;a) =1$ and $\gamma_i =0$ for this model. This means the only modification comes in the form of the time-dependent $A(a)$. 

Similar models starting from a Lagrangian \cite{Pourtsidou:2013nha} have also been derived. These so called Type 3 models also involve no background energy exchange and are interesting in their ability to suppress late-time linear growth, in doing so alleviate the CMB-LSS $\sigma_8$ discrepancy. These Type 3 models  predict three extra terms in Eq.(\ref{eq:Perturb2})  proportional to  $\theta$, the dark energy velocity divergence $\theta_{\rm DE}$ and the dark energy density contrast $\delta_{\rm DE}$. The latter terms are absent in the DS models and as shown in \cite{Skordis:2015yra} there is no obvious way to remove the last contribution without removing the interaction all-together in Type 3 models. Despite this, the DS and Type 3 models should be qualitatively similar in their predictions \cite{Baldi:2016zom}. In principle the extension to include these terms in Eq.(\ref{eq:Perturb2}) is one of straightforward derivation.

\subsection{Numerical PT treatment}  
\label{subsec:numerical_method}
Once we have specified $A(a)$, $\mu(k;a)$ and $\gamma_i$ (for $i \in \{ 2,3,4 \}$) we can employ the algorithm described in \cite{Taruya:2016jdt} to calculate the perturbative kernels, $F_n(\bfk_1, \dots, \bfk_n;a)$ for $n\leq 4$. We outline this method here.  Using Eq.(\ref{eq:Perturb1}) and Eq.(\ref{eq:Perturb2}) and the field definitions  given in Eq.(\ref{nth1}) and Eq.(\ref{nth2}) we have the following coupled set of evolution equations for the $n^{th}$ order kernels  
\begin{align}
\frac{\partial F_n(\bfk_1, \dots, \bfk_n;a)}{\partial a}  =& -\frac{1}{a} \left[ G_n(\bfk_1, \dots, \bfk_n;a) + \sum_{j=1}^{n-1} \alpha(\bfk_{1 \cdots j},\bfk_{j+1 \cdots n};a) G_j(\bfk_1,\cdots,\bfk_j;a) F_{n-j}(\bfk_{j+1}, \cdots, \bfk_n;a) \right], \label{evoleqn:fn} \\
\frac{\partial G_n(\bfk_1, \dots, \bfk_n;a)}{\partial a}  =& -\frac{1}{a} \Bigg[ \left(2+ A(a) + \frac{a H'}{H}\right) G_n(\bfk_1, \dots, \bfk_n;a) + \frac{3 \Omega_{m,0} H_0^2}{2 a^3 H^2} {\mu(k;a)} F_n(\bfk_1, \dots, \bfk_n;a)  \nonumber \\ &  \qquad  +  \frac{1}{2} \sum_{j=1}^{n-1} \beta(\bfk_{1 \cdots j},\bfk_{j+1 \cdots n}) G_j(\bfk_1,\cdots,\bfk_j;a) G_{n-j}(\bfk_{j+1}, \cdots, \bfk_n;a)  + S_{k,n}(\bfk_1, \cdots, \bfk_n;a) \Bigg], \label{evoleqn:gn}
\end{align}
where $k = |\bfk_1 + \dots \bfk_n|$ and $S_{k,n}$ is the $n^{th}$ order source function. This has been employed up to 3rd order in  \cite{Taruya:2016jdt,Bose:2016qun} for the one-loop power spectrum calculation. For the one-loop bispectrum calculation this must be specified up to 4th order which we present below 
\begin{align}
 S_{k,4}(\bfk_1, \bfk_2,\bfk_3, \bfk_4;a) = & \sum_{j=1}^{3} \gamma_2(\bfk_{1 \cdots j},\bfk_{j+1 \cdots 4}) F_j(\bfk_1,\cdots,\bfk_j;a) F_{4-j}(\bfk_{j+1}, \cdots, \bfk_4;a) \nonumber \\ 
 &+ \gamma_3(\bfk_1,\bfk_2,\bfk_{34};a) F_1(k_1;a)F_1(k_2;a)F_2(\bfk_3,\bfk_4;a)\nonumber \\ 
 &+  \gamma_4(\bfk_1,\bfk_2,\bfk_3,\bfk_4;a) F_1(k_1;a)F_1(k_2;a)F_1(k_3;a)F_1(k_4;a).
 \end{align}
A final complication is that we require the symmetrised kernels to construct the spectra  
\begin{align}
F_n^{sym}(\bfk_1,\cdots,\bfk_n;a) &= \frac{1}{n!} \left[ F_n(\bfk_1,\cdots,\bfk_n;a) + \mbox{perms} \right],\\
G_n^{sym}(\bfk_1,\cdots,\bfk_n;a) &= \frac{1}{n!} \left[ G_n(\bfk_1,\cdots,\bfk_n;a) + \mbox{perms} \right],
\end{align}
where the permutations are over the wave vector arguments. We simply include the relevant permutations on the right hand sides of Eq.(\ref{evoleqn:fn}) and Eq.(\ref{evoleqn:gn}) so that we solve for the symmetrised kernels.

Given Einstein-de Sitter initial conditions (valid at early times during matter domination), the coupled set of differential equations can be solved for $F_n$ and $G_n$ once we know $F_i$ and $G_i$ for $i<n$. We solve for these order by order. In practice, all coupled sets of equations for $F_i$ and $G_i$ with $i\in \{1,2,3,4\}$ are solved for simultaneously as one large set. Further, this set must be solved for each desired combination of wave vectors in $F_n(\bfk_1, \cdots, \bfk_n;a)$, specifically those combinations appearing in the expressions Eq.(\ref{tree1}) to Eq.(\ref{fourtho}). For example, the $F_4(\bfk_i,\bfk_j,\bfp,-\bfp;a)$ kernels needed in Eq.(\ref{fourtho})  depend on terms involving specific 3rd, 2nd and 1st order kernels which also need to be solved for. In the end, although Eq.(\ref{tree1}) to Eq.(\ref{fourtho}) only explicitly depend on 28 specific kernels; 3 $\times$ 1st order, 13 $\times$ 2nd order, 9 $\times$ 3rd order and 3 $\times$ 4th order, each for a specific combination of wave vectors, we must solve a total of 47 coupled sets numerically; 8$\times$ 1st order, 21 $\times$ 2nd order, 15$\times$  3rd order  and 3 $\times$ 4th order. We solve these sets using the {\tt gsl} package {\bf odeiv2} with a Runge-Kutta Prince-Dormand (8,9) method.

The kernels are then integrated over wave vector magnitude and 2 angular variables and so this large set of differential equations must be solved per integration step for each of the  3 integrals. This results in a large number of calls to the differential equation solver which can be very time costly depending on the accuracy demands. Further, as we do not implement a fully IR-Safe integral \cite{Baldauf:2014qfa} in the above expressions, the numerical accuracy of the differential equation solver and loop integration routine should be carefully tuned so as to balance time cost and numerical accuracy. In the case of the one-loop bispectrum, the level of numerical accuracy need not be as high as the power spectrum given the larger statistical errors in current and upcoming surveys. In general, for our results in the next section, the average time cost of producing 20 equilateral shape one-loop bispectrum points between $k=0.001 - 0.3  h \mbox{Mpc}^{-1}$ is 350 seconds. This varies over model and slightly over redshift with the $f(R)$ model taking the longest. We give more details on numerical accuracy and time costs in Appendix A. 


\section{Comparison with $N$-body simulations} 
\label{sec:comparison}

In this section,  we compare our numerical PT predictions with results from cosmological simulations, specifically paying attention to the bispectrum in the three representative models described in Sec.~\ref{subsec:models} as well as $\Lambda$CDM (i.e. GR). Also, the validity of several approximations are tested against the full numerical PT treatment and $N$-body measurements.

\subsection{$N$-body simulations}

In this paper, we use a Comoving Lagrangian Acceleration (COLA)  \cite{Tassev:2013pn,Howlett:2015hfa} code to create the simulation data for nDGP, Hu-Sawicki $f(R)$ gravity and $\Lambda$CDM (GR). To be precise, we use the modified MG-PICOLA code described in Ref.~\cite{Winther:2017jof}. Each simulations uses a cubic box of side length $1024 \mbox{Mpc} h^{-1} $ and $1024^3$ particles and we employ $20$ independent such realisations \footnote{Initial seeds used to create the $20$ realisations are the same among all three models.} starting from $z=49$ with initial conditions generated by second-order Lagrangian PT. We adopt the initial power spectrum determined by WMAP9 \cite{Hinshaw:2012aka}: $\Omega_{\rm m} = 0.281$, $\Omega_{\rm b} = 0.046$, $h=0.697$, $n_{\rm s}=0.971$ and $\sigma_8=0.844$. In nDGP and $f(R)$ gravity models, one also needs to specify one more free parameter, for which we set $\Omega_{\rm rc}=0.438$ and $|\bar{f}_{R0}| = 10^{-4}$, respectively. Although these values have already been ruled out by observations (e.g. \cite{Barreira:2016ovx,Burrage:2017qrf}), they are still useful to see if our PT predictions properly describe the non-linear effects of MG. 

We use a single realisation for the DS model. Specifically, the data set is taken from Ref.~\cite{Baldi:2016zom}. The simulation was created with a modified version of {\tt GADGET-2} \cite{Springel:2005mi} that consistently implements the effects of the momentum exchange between dark matter particles and a homogeneous dark energy. This data assumes the equation-of-state parameter for dark energy $w=-1.1$ and the interaction parameter of $\xi = 10$ bn$\mbox{GeV}^{-1}$ [see Eq.(\ref{interactionterm}) in Sec.II B]. The output redshifts of the data, the box size, and number of dark matter particles are the same as in the three models described above. On the other hand, the initial power spectrum uses a slightly different set of parameters: $\Omega_{\rm m} = 0.308$, $\Omega_{\rm b} = 0.0482$, $h=0.678$, $n_{\rm s}=0.966$ and $\sigma_8 = 0.852$. For more detailed information, we refer the readers to Ref.~\cite{Baldi:2016zom}.  

We measure the power spectrum and bispectrum at $z=0$, $z=0.5$ and $z=1$ from the grid-assigned density field using a Fast-Fourier Transform based estimator (e.g. \cite{Baldauf:2014qfa,Sefusatti:2015aex}). We use the cloud-in-cells interpolation for the density assignment of particles onto a $512^3$ mesh and correct the window function. In what follows, except for the DS model, we present the measured results of power spectrum and bispectrum averaged over the $20$ realisations, with quoted error bars determined by twice the standard error of the mean (Figs.~\ref{lcdma}-\ref{fra}). For the DS model, shown in Fig.~\ref{dsb}, we use only a single-realisation, and do not quote the error bars.

\subsection{Testing numerical PT predictions}

We first investigate the numerical PT's performance. Figs.~\ref{lcdma}-\ref{dsb} summarize the measurements and predictions of the bispectrum in equilateral (middle) and isosceles (right, with fixed wavenumber $k_1=k_2=0.096\,h$$\mbox{Mpc}^{-1}$) configurations in different cosmologies, plotted as a function of wavenumber $k\equiv k_1=k_2=k_3$ and the angle defined by $\theta=\cos^{-1}(\hat{\bfk}_1\cdot \hat{\bfk}_2)$, respectively. The results of the power spectrum are also presented on the left. Note that all the results are multiplied by $k^{3/2}$ ($k^3$) for power spectrum (bispectrum).

In each of the three figures, the top panels compare the results of the $N$-body simulations (red crosses) with numerical PT predictions at tree (red dashed) and one-loop (green solid) order. Clearly, the one-loop predictions better describe the non-linear enhancement of clustering amplitudes at all redshifts, although the agreement with simulation results is restricted to a narrow range of wavenumbers in the power spectrum. This is a well known problem of SPT \cite{Carlson:2009it}, and explains why there has been various techniques developed to improve the SPT prediction. By contrast, the performance of the one-loop bispectrum is much better, with the predictions agreeing well with simulations at $z\gtrsim0.5$ for a rather wide range of $k$ (i.e. $k\lesssim0.3\,h$$\mbox{Mpc}^{-1}$). The exception is  the $f(R)$ gravity model, where a rather strong enhancement of the bispectrum is seen at small scales. The discrepancy at $k\gtrsim0.15\,h$$\mbox{Mpc}^{-1}$ is associated with our setup of simulation parameter, $|\bar{f}_{R0}| = 10^{-4}$, with which 
the screening mechanism is ineffective, and gravity becomes stronger at small scales. We anticipate that the one-loop prediction in $f(R)$ gravity also reproduces the simulations as well at the other models for a reasonable choice of $|\bar{f}_{R0}|$ with which the chameleon screening can work.

To see the impact of non-linear growth, the middle panels show the ratio of measurements and one-loop SPT to the tree level theory predictions. Despite the fact that the bispectrum receives rather large non-linear corrections compared to the power spectrum, the one-loop bispectrum reasonably explains the $N$-body trends. Further, for $\Lambda$CDM, nDGP and DS models, we plot the predictions of a non-linear fitting formula in both top and middle panels, depicted as blue solid lines. The results shown in the power spectrum (left) are obtained from the revised version of {\tt halofit} \cite{Smith:2002dz} by Ref.~\cite{Takahashi:2012em}. To be strict, {\tt halofit} can apply only to the GR case, but it has been frequently used in MG models close to $\Lambda$CDM in the literature. We thus similarly use it to predict the non-linear power spectrum based on the linear theory prediction in each model. On the other hand, we use the fitting formula for the bispectrum given by Ref.~\cite{GilMarin:2011ik} (see also \cite{Scoccimarro:2000ee}). The prediction of the non-linear bispectrum is based on the non-linear power spectrum and a modified second-order PT kernel that is calibrated with $N$-body simulations. This is again valid only in GR. To apply it to non-standard models, we follow the treatment proposed in Ref.~\cite{Namikawa:2018erh}, and slightly modify the calibrated second-order kernel so as to consistently recover the tree-level SPT results at large scales. In Appendix \ref{appendix:fitting_formula_bispec}, we present the explicit expression for the fitting formula, and briefly mention how to specifically implement it in each model. Note we do not include such non-linear fitting formula predictions for $f(R)$. Such a formula is non-trivial due to scale-dependent growth. For the power spectrum, we refer the interested reader to a proposed and tested extension of the {\tt halofit} approach to $f(R)$ in Ref.~\cite{Zhao:2013dza}. Such an extension for the bispectrum has not been studied. 

Overall, the fitting formula reproduces the simulation results quantitatively well. This is indeed true for the power spectrum. A closer look at the bispectrum, however, reveals that the fitting formula tends to slightly under predict the amplitude at $z\gtrsim0.5$. Rather, at  scales of $k\lesssim0.3\,h$$\mbox{Mpc}^{-1}$, the simulation results are better described by the one-loop SPT predictions, and this is not only in $\Lambda$CDM, but also in the nDGP and DS models. Although the differences are not large, the discrepancy between the fitting formula and simulation may be partly ascribed to an improper treatment of the non-linearity in the fitting formula associated with the modification of structure formation. Another issue may be the difference in cosmology used in this paper from that used to calibrate the fitting formula in \cite{GilMarin:2011ik}. As the cosmology dependence of the general formula has been shown to be weak \cite{Scoccimarro:2000ee} and that our cosmology is similar to that used to calibrate the formula we don't expect this to be a large effect. We do note that our value of $\sigma_8$ is $\sim 6\%$ larger than \cite{GilMarin:2011ik} which introduces more non-linearity and so may play a small role in the fitting formula performance, but we also expect these effects to play a less significant role at the scales considered in this paper. There might also be possible systematics in our $N$-body simulations as the accuracy of the COLA and MG-PICOLA code have not yet been tested for higher-order statistics. Nevertheless, one encouraging point is that the one-loop bispectrum provides a rather accurate prediction at $z\gtrsim0.5$, comparable to the fitting formula, and can be used for a quantitative comparison with observations.

\subsection{Comparison with analytic PT treatment} 

In a limited class of generalised cosmological models,  analytic PT calculations are known to become tractable. The analytic PT kernels are very useful in that statistical predictions are quickly and efficiently calculated. nDGP, DS and $\Lambda$CDM are such  models. In particular, assuming the Einstein-de Sitter (EdS) universe, the time dependence is analytically expressed in terms of the scale factor in GR. Then, one often invokes the approximation that the analytic calculations made in the EdS universe are generalised to non-EdS models by simply replacing the scale factor with the linear growth factor, $F_1(a)$, obtained in the non-EdS model. This is the so-called EdS approximation, and has been frequently  used in the literature as a reasonably accurate approximation in $\Lambda$CDM and its variants. This is indeed true and has been tested in the power spectrum case (for example \cite{Takahashi:2008yk,Hiramatsu:2009ki,Fasiello:2016qpn,Bose:2018orj}), but its validity to the bispectrum calculation has not been thoroughly tested, especially for  models beyond $\Lambda$CDM. Further, another simplification that one can apply to generalised cosmologies is to just take into account the linear-order modification to gravity, ignoring all  non-linear modifications. To be precise, in our basic equations, this amounts to retaining $\mu(k;a)$ while setting $\gamma_i=0$, and $A(a)=0$ \footnote{Setting $A(a)=0$ also changes the linear growth, but we will normalise with the linear predictions in this case to highlight only non-linear effects.}. We call this the un-screened approximation (UsA), and critically examine the validity of this treatment to the bispectrum. 

\begin{figure}[H]
\centering
  \includegraphics[width=17cm,height=6.5cm]{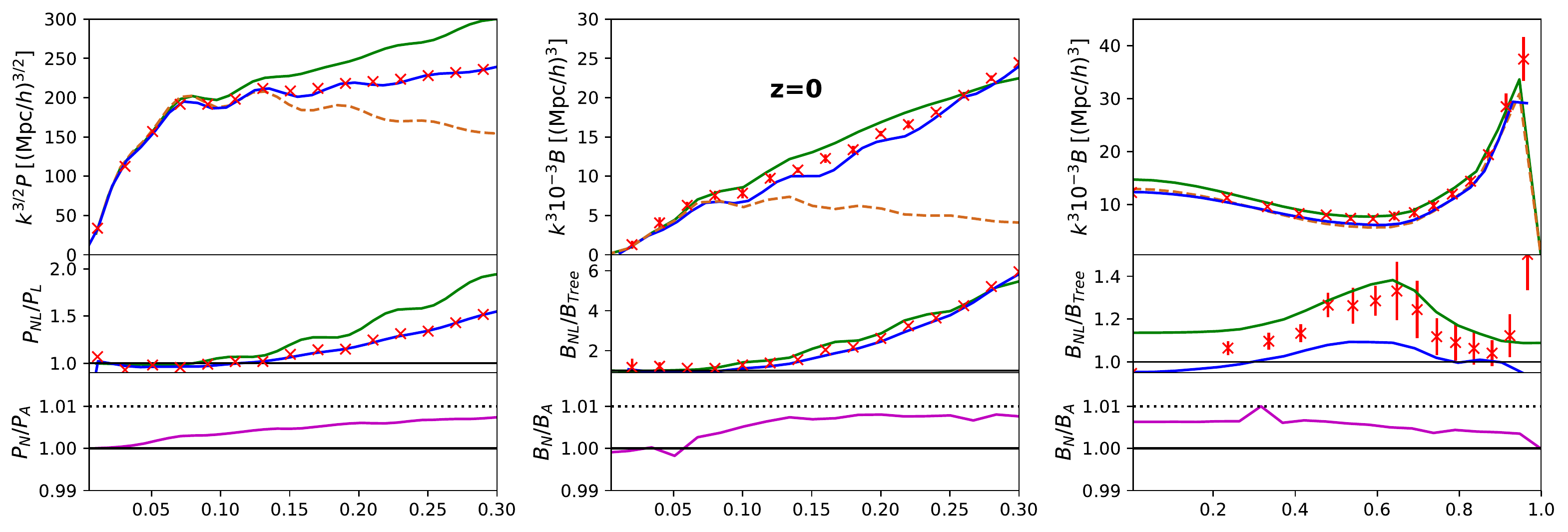} \\ 
    \includegraphics[width=17cm,height=6.5cm]{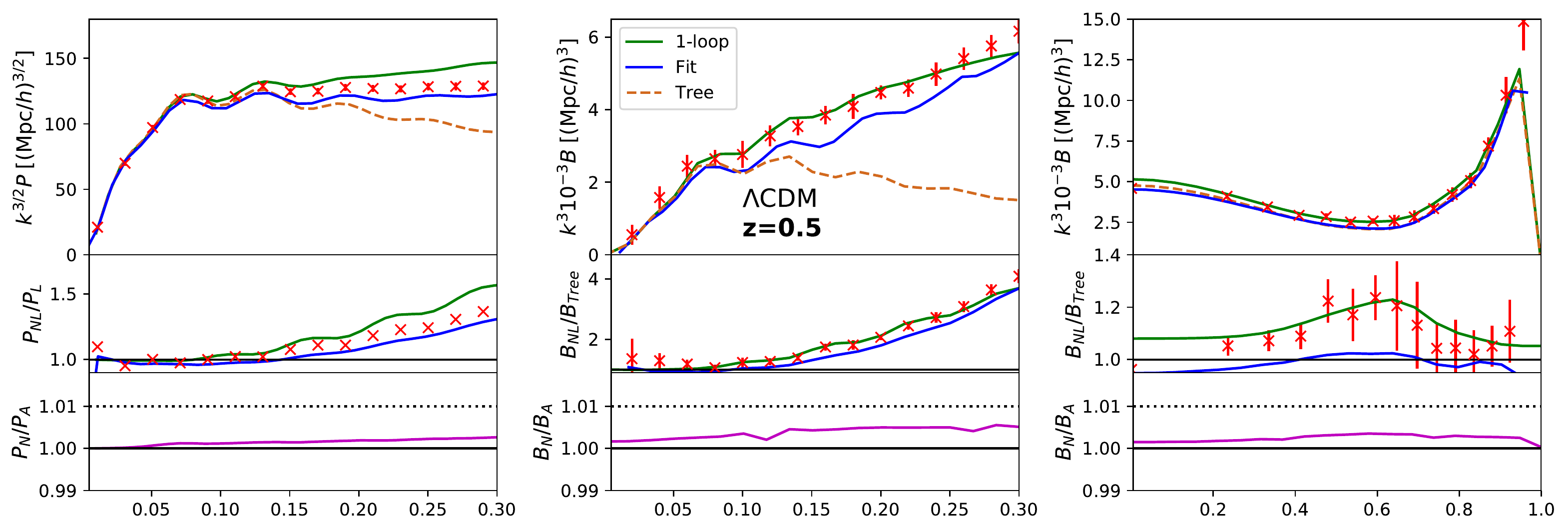} \\ 
      \includegraphics[width=17cm,height=6.5cm]{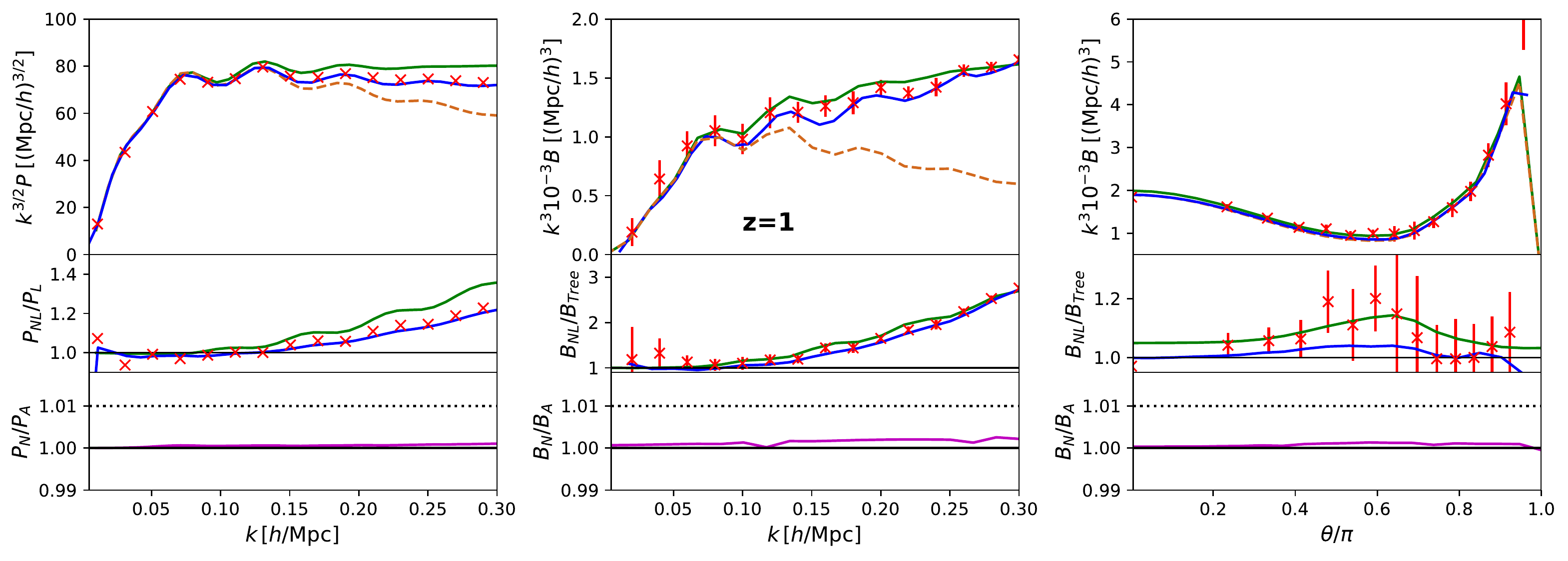}
  \caption[CONVERGENCE ]{Comparison of the numerical PT treatment with $N$-body simulation in $\Lambda$CDM model (GR). Left panels show the power spectrum, while middle and right panels plot the bispectrum, particularly showing the scale dependence of the equilateral shape ($k\equiv k_1=k_2=k_3$) as function of $k$, and the configuration dependence of the isosceles shape at $k_1=k_2=0.096\,h$\,Mpc$^{-1}$, plotted as a function of $\theta\equiv \cos^{-1}(\hat{\bfk}_1\cdot\hat{\bfk}_2)$. From top to bottom, the results at $z=0$, $0.5$, and $1$ are summarized. In each panel, top panels compare the $N$-body results (red crosses) with numerical PT predictions at tree-level (red dashed) and one-loop (green solid) order. As a reference, the results from fitting formula (see text for details) are also plotted (blue solid). The quoted error in simulations is twice the standard error of the mean over the $20$ realizations. Note that all the results of measurement and prediction are multiplied by $k^{3/2}$ for power spectrum and $k^3$ for bispectrum. The second panels from the top present the ratio of non-linear predictions and measurements to the linear theory or tree-level PT prediction, i.e., $P_{\rm NL}/P_{\rm L}$ (left) and $B_{\rm NL}/B_{\rm tree}$ (middle and right). Meanings of the symbols and line types are the same as in the top panels. Finally, the bottom panels compare the analytic PT predictions with numerical PT results. What is plotted is the ratio of numerical to analytic PT results adopting EdS approximation }
  \label{lcdma}
\end{figure}

\begin{figure}[H]
\centering
  \includegraphics[width=17cm,height=6.5cm]{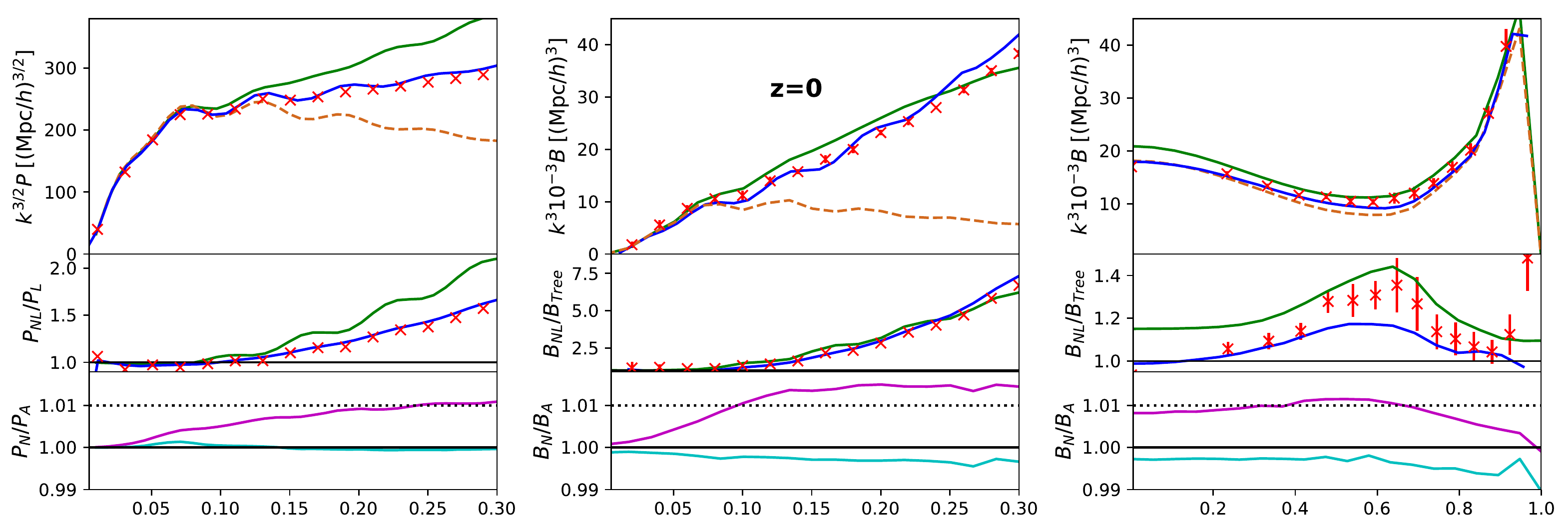} \\
   \includegraphics[width=17cm,height=6.5cm]{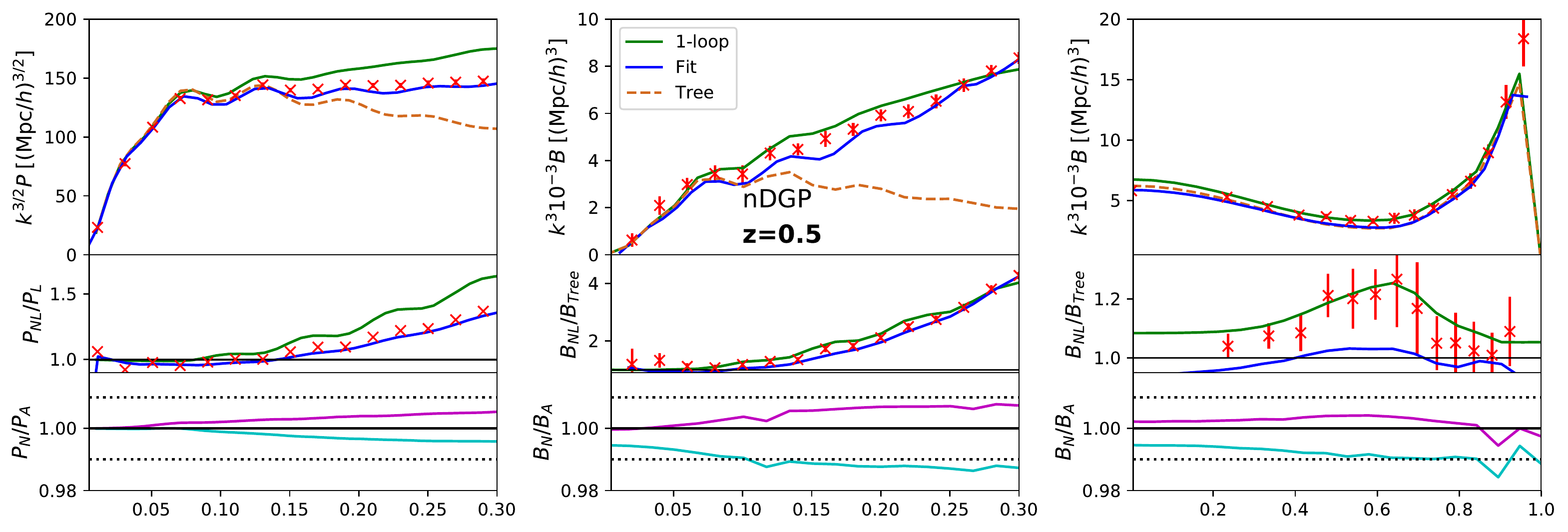} \\ 
    \includegraphics[width=17cm,height=6.5cm]{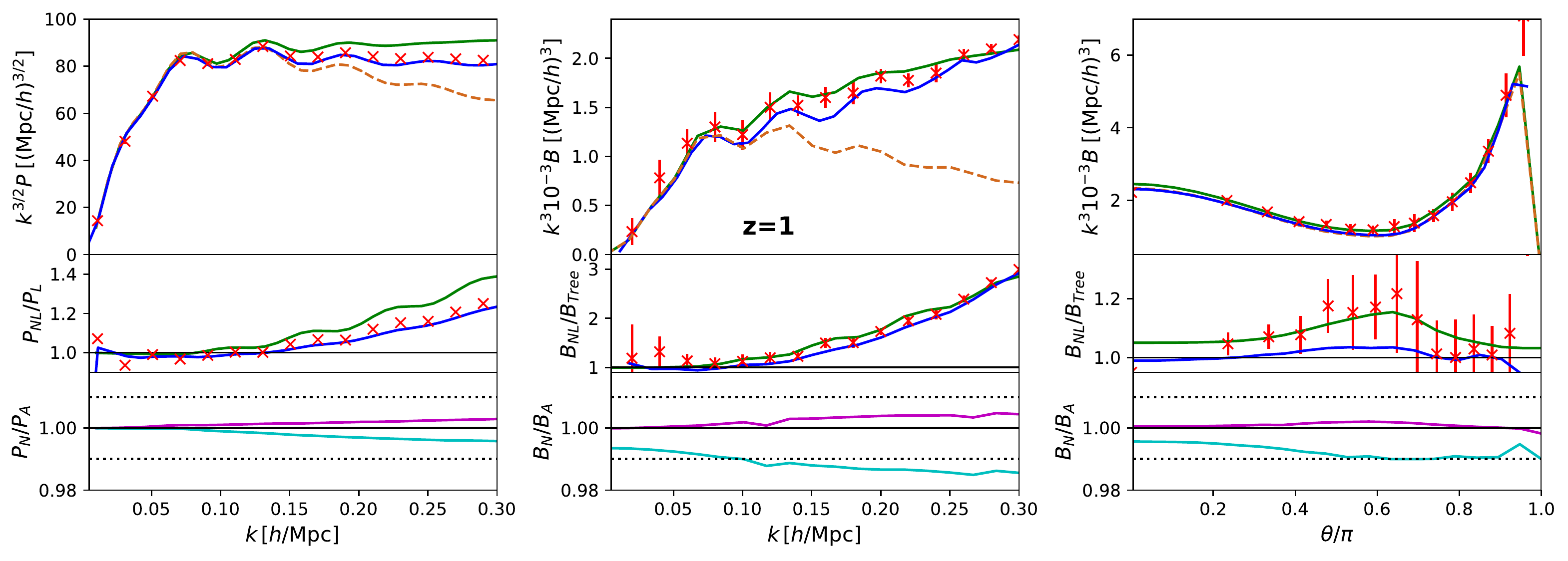}
  \caption[CONVERGENCE ]{Same as in Fig.~\ref{lcdma}, but the results in the nDGP model with $\Omega_{\rm rc}=0.438$ are shown. In the bottom panels we also plot cyan lines which are the ratio of the one-loop numerical PT to analytic PT predictions adopting both the EdS and UsA.}
\label{dgpa}
\end{figure}

\begin{figure}[H]
\centering
  \includegraphics[width=17cm,height=6.5cm]{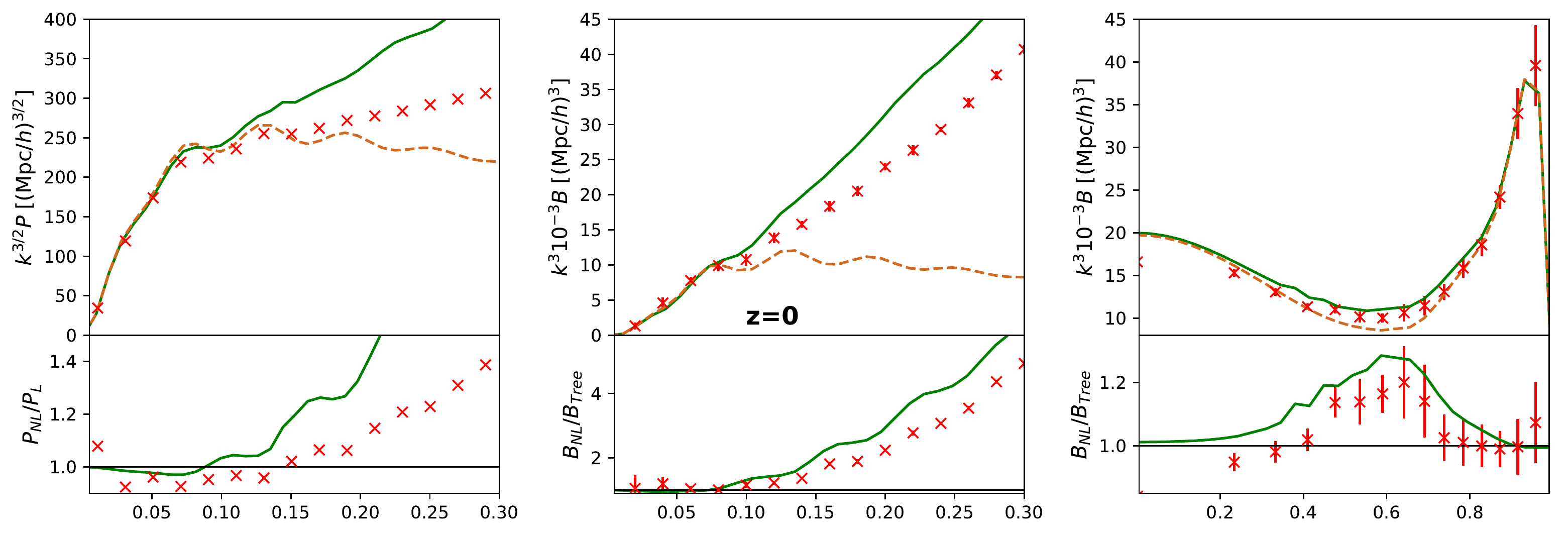} \\ 
   \includegraphics[width=17cm,height=6.5cm]{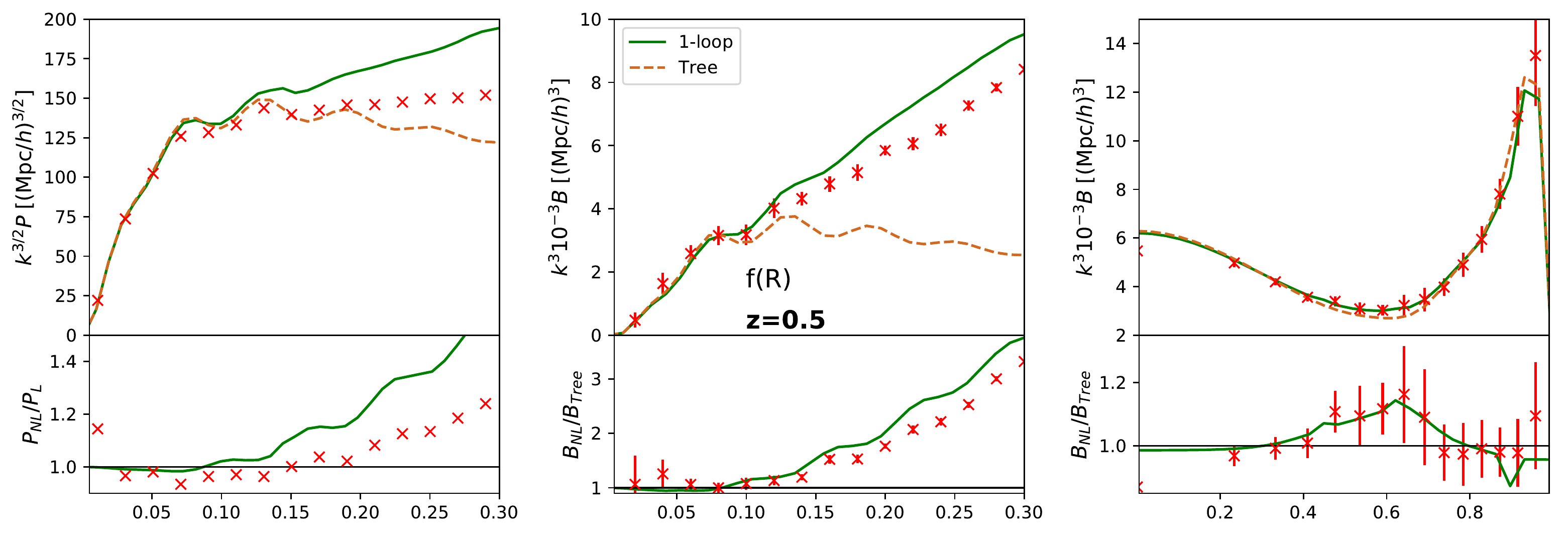} \\ 
    \includegraphics[width=17cm,height=6.5cm]{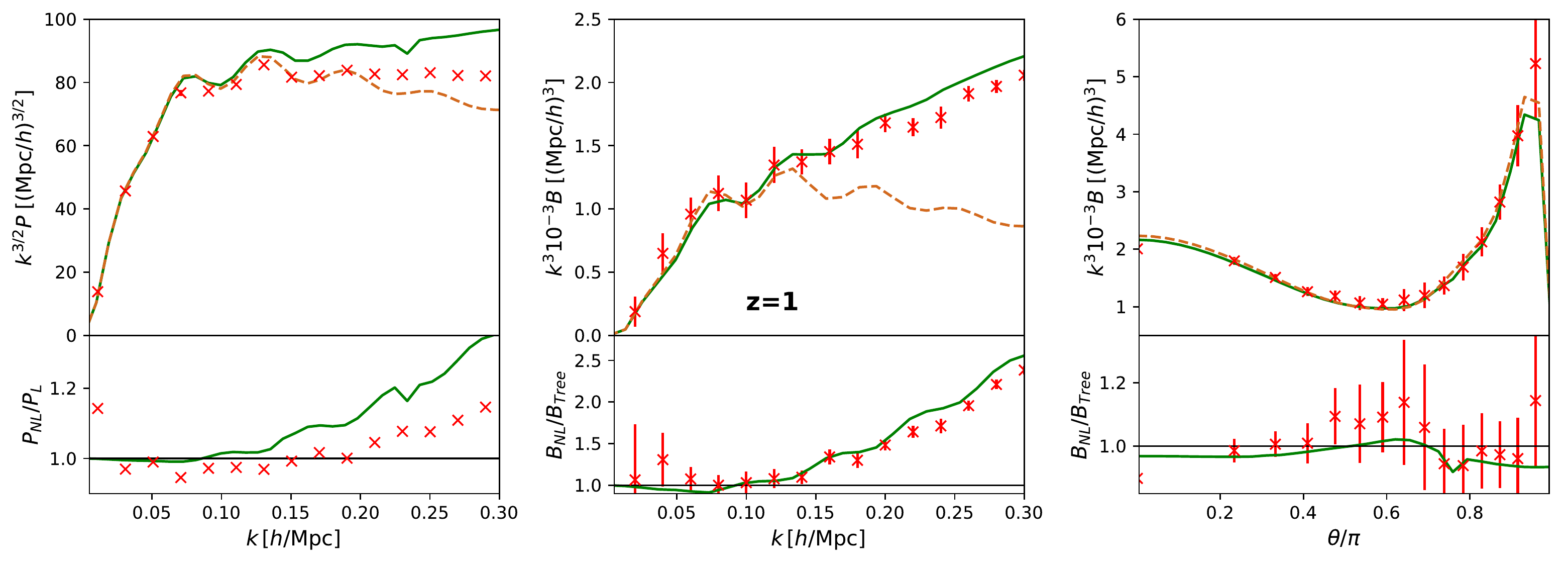}
  \caption[CONVERGENCE ]{Same as in Fig.~\ref{dgpa}, but the results in $f(R)$ gravity with model parameter $|\bar{f}_{R0}| = 10^{-4}$ are shown. We  do not plot a fitting formula in this case. Also, since an analytic PT treatment is intractable in this model, we did not make a comparison between analytic and one-loop numerical PT predictions in the bottom panels. }
\label{fra}
\end{figure}

\begin{figure}[H]
\centering
  \includegraphics[width=17cm,height=6.5cm]{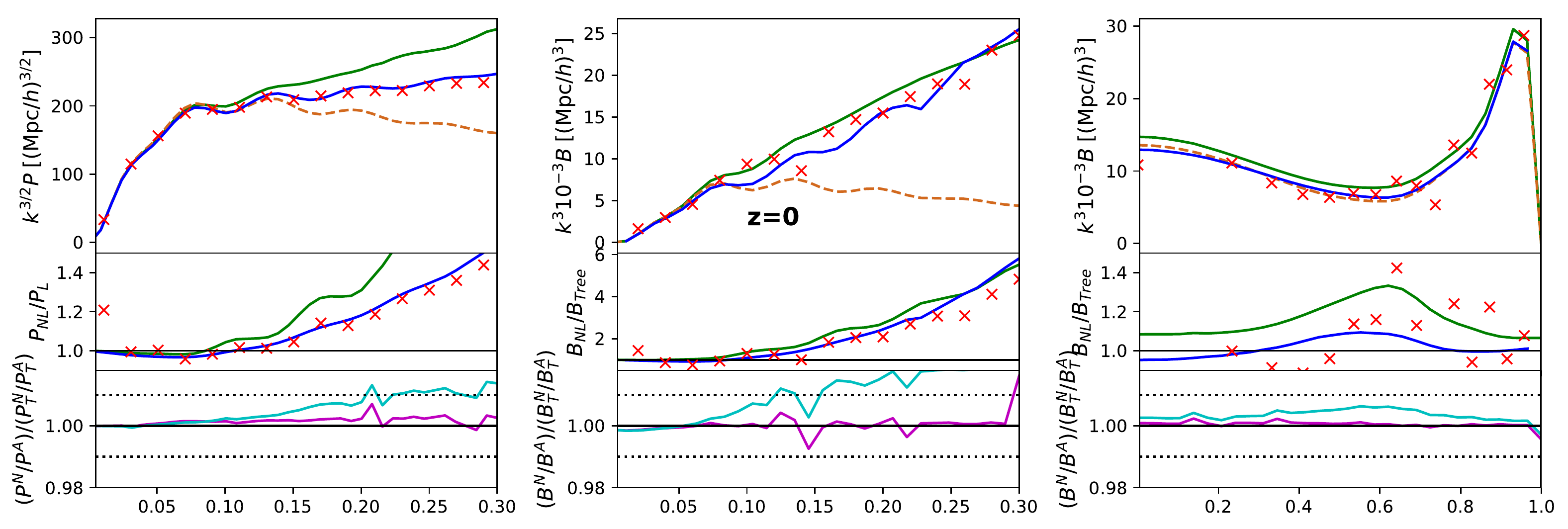} \\
   \includegraphics[width=17cm,height=6.5cm]{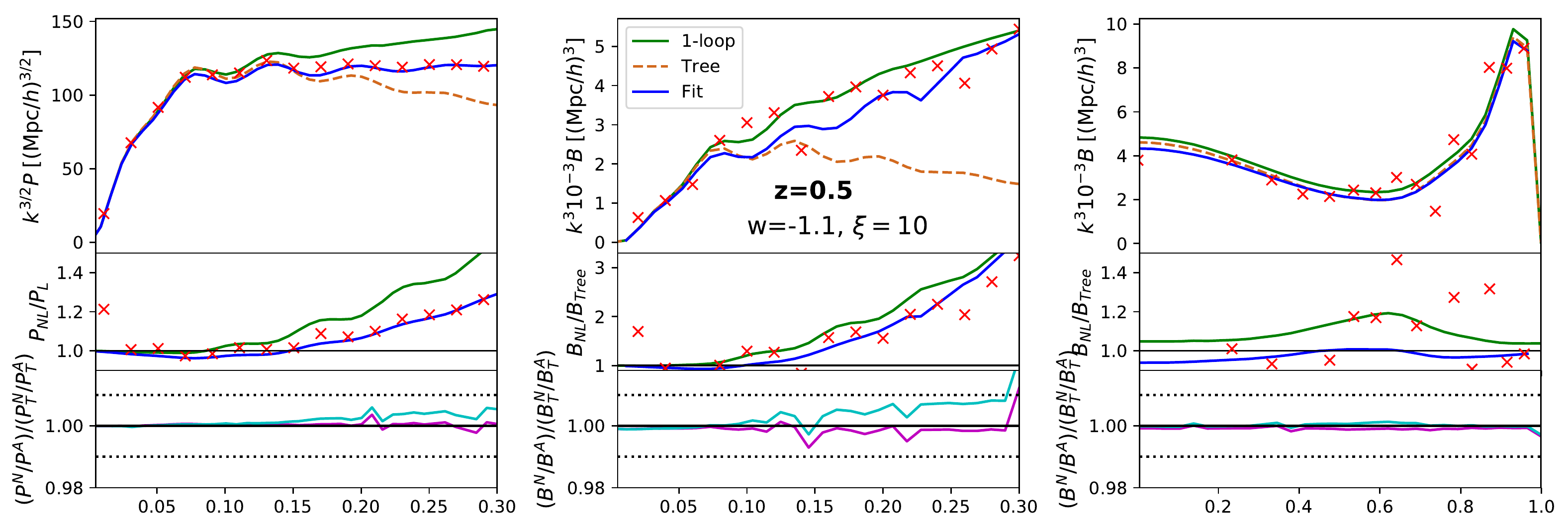} \\ 
    \includegraphics[width=17cm,height=6.5cm]{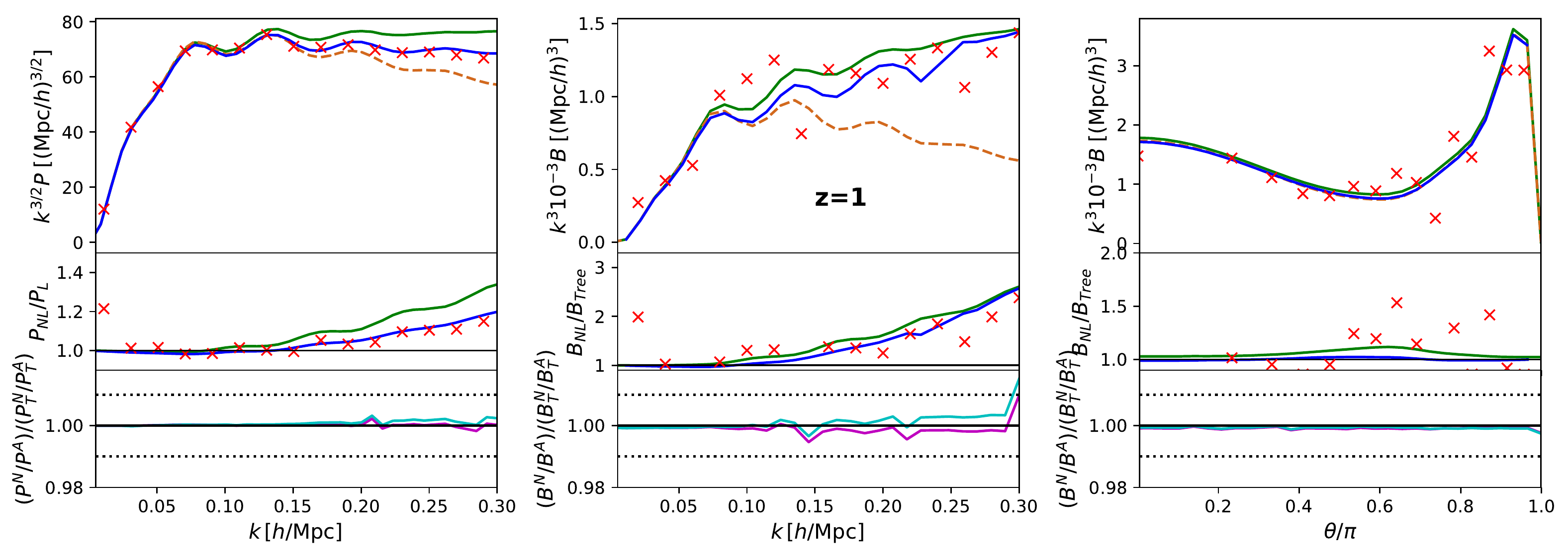}
  \caption[CONVERGENCE ]{Same as in Fig.~\ref{dgpa}, but the results in the DS model with interaction parameter $\xi=10$ are shown. In the bottom panels we show the ratio $P_{\rm N}/P_{\rm A}$ and $B_{\rm N}/B_{\rm A}$ and further divide these ratios by the same ratio at tree-level order so as to asymptotically approach unity in the limit $k\to0$, i.e. $(P_{\rm N}/P_{\rm A})/(P_{\rm N,T}/P_{\rm A,T})$ (left) and $(B_{\rm N}/B_{\rm A})/(B_{\rm N,T}/B_{\rm A,T})$.}
\label{dsb}
\end{figure}

In the bottom panels of Figs.~\ref{lcdma}, \ref{dgpa}, and \ref{dsb},  we compare the analytic PT treatment with the numerical PT prediction. What is shown here is the ratio of numerical PT results ($P_{\rm N}$ or $B_{\rm N}$) to the analytic PT results based on the EdS or UsA ($P_{\rm A}$ or $B_{\rm A}$), i.e. $P_{\rm N}/P_{\rm A}$ for the power spectrum and $B_{\rm N}/B_{\rm A}$ for bispectrum. Solid magenta lines are the results adopting the EdS approximation, while cyan lines, shown in Figs.~\ref{dgpa} and \ref{dsb}, represent the cases adopting both the EdS and UsA. For the nDGP model, the analytic expressions for the PT kernels are presented in Ref.~\cite{Koyama:2009me} up to the third order, and we use them for the one-loop calculation of the power spectrum. To compute the one-loop bispectrum, we further need the fourth-order PT kernel which we have derived in this paper, presented in Appendix \ref{appendix:kernel_F4_in_DGP}. Also, for the DS model, the ratios, $P_{\rm N}/P_{\rm A}$ and $B_{\rm N}/B_{\rm A}$ are further divided by those at tree-level order, so as to asymptotically approach unity in the limit $k\to0$. 

In most of the cases, both the EdS and UsA produce an error at sub-percent level within the validity range of one-loop SPT predictions, roughly $k\lesssim0.08-0.15\,h$$\mbox{Mpc}^{-1}$ for power spectrum and $k\lesssim0.1-0.3\,h$$\mbox{Mpc}^{-1}$ for bispectrum at the redshift range $z=0-1$ \footnote{This range is here  determined by comparing with the mean and twice the standard error of the measurements.}. This is indeed the case for the power spectrum in all models, but a closer look at the bispectrum reveals that the error relative to the numerical PT results is more pronounced. In particular, in the nDGP model, this systematic error can reach the percent level, and at higher redshifts $z\gtrsim0.5$, the combination of the EdS and UsA (cyan curve) produces an even larger systematic error. Even employing only the EdS approximation is potentially problematic at $z=0$.  In the power spectrum case, this level of deviation is shown to be an issue in constraining MG theories in stage-IV spectroscopic surveys \cite{Bose:2016qun}. Although statistical error of the bispectrum would be certainly larger even for such a survey, combining all possible triangular configurations may accumulate the systematics, potentially leading to a biased constraint. A deeper study into this would be an important subject for practical applications of the one-loop bispectrum to data.

\section{Discussion: impact of non-linear MG effects on the bispectrum}
\begin{figure}[H]
\centering
  \includegraphics[width=18cm,height=6.25cm]{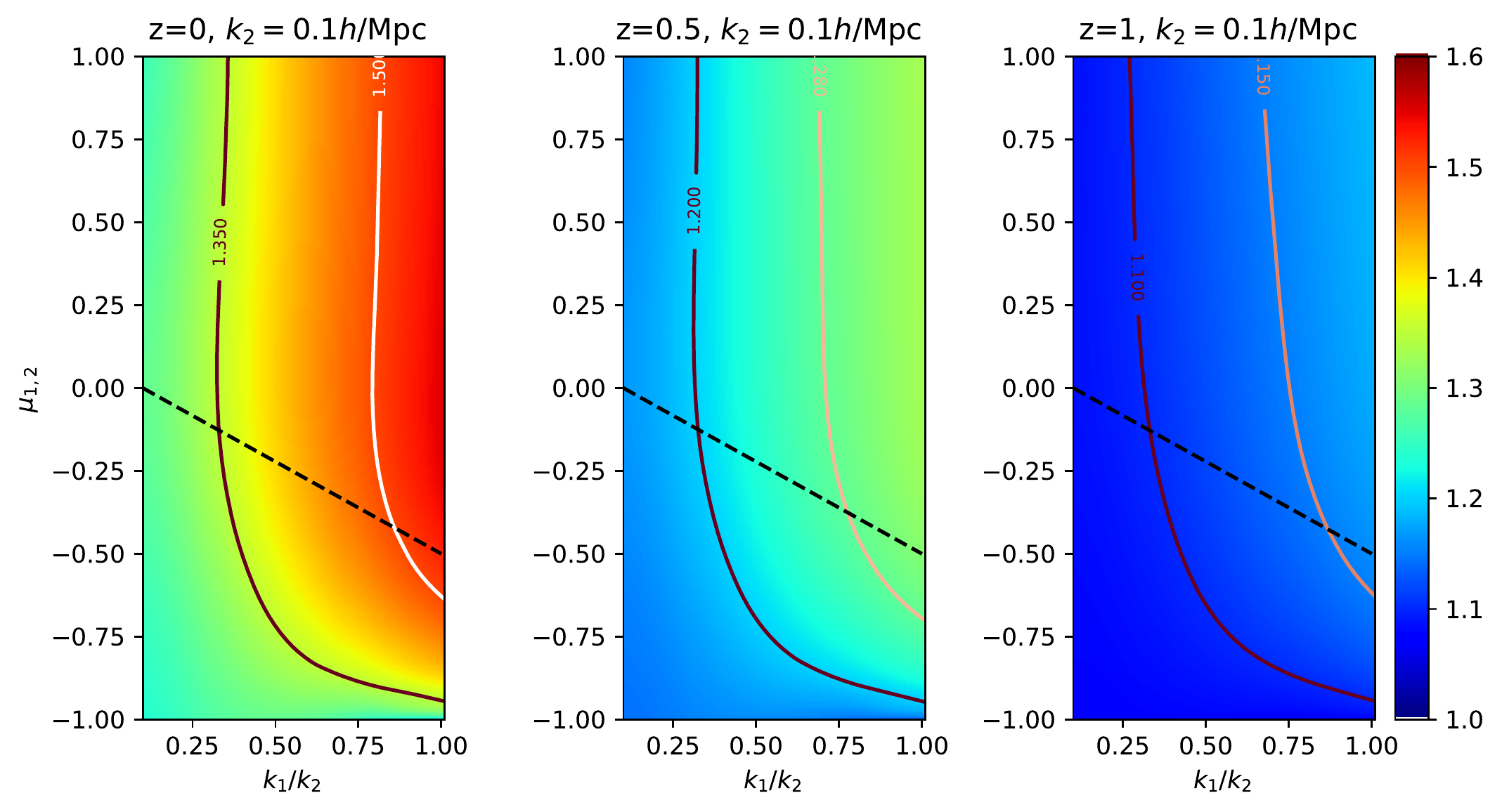}
  \includegraphics[width=18cm,height=6.25cm]{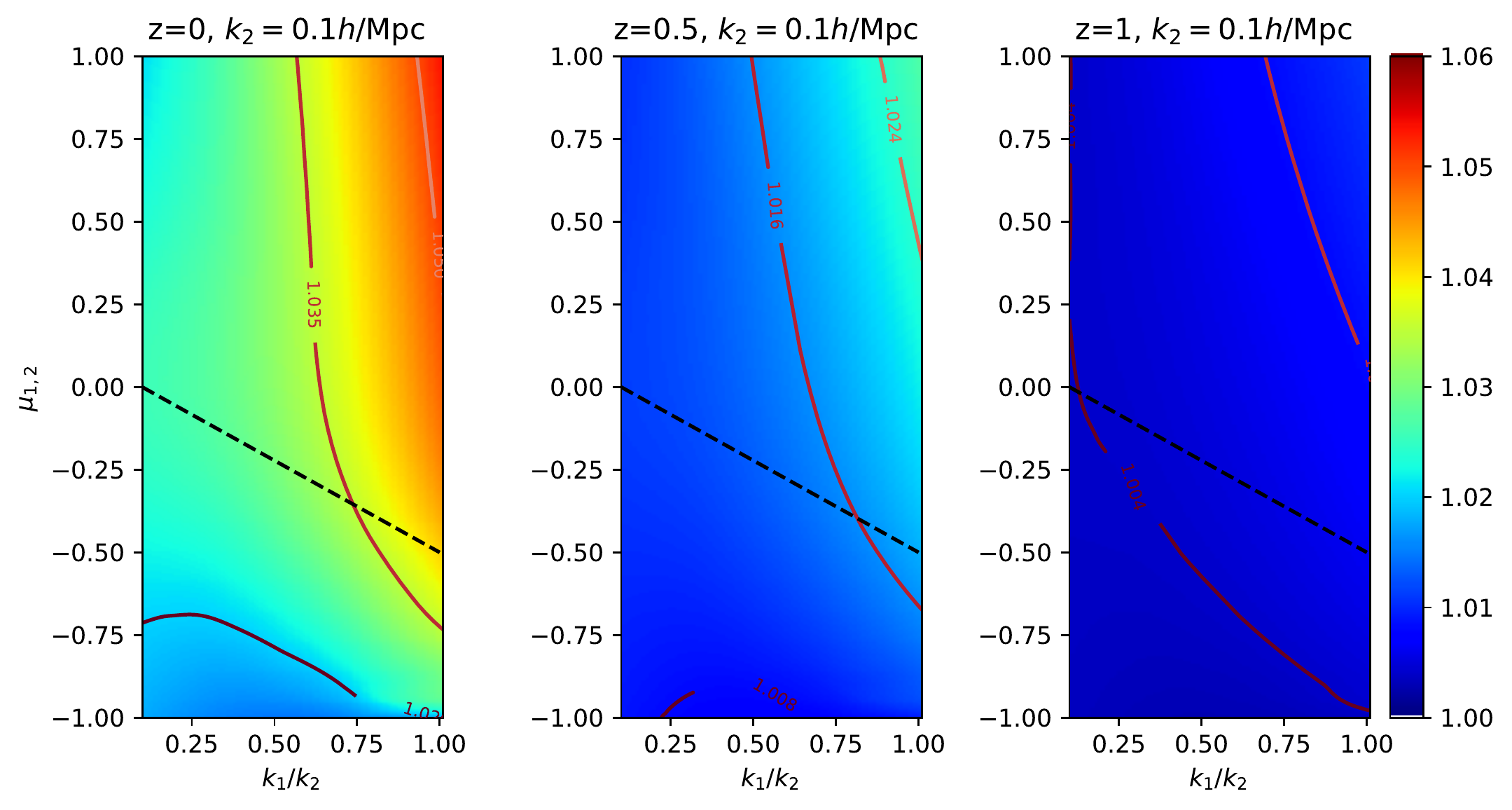}
  \caption[CONVERGENCE ]{Ratio of tree-level SPT prediction of the bispectrum in $f(R)$ gravity model to that in GR, plotted as function of $k_1/k_2$ and $\mu_{1,2} = (\hat{\bfk}_1 \cdot \hat{\bfk}_2)$, fixing $k_2$ to $0.1\,h\,$$\mbox{Mpc}^{-1}$. {\bf Top} and {\bf bottom} panels show the results with model parameter $|\bar{f}_{R0}| = 1 \times 10^{-4}$ and  $2.5\times10^{-6}$, respectively. The results are shown at $z=0$ (left), $0.5$ (middle), and $1$ (right). The black dashed lines mark the equilateral shape. }
\label{cont3}
\vspace*{0.5cm}
\centering
  \includegraphics[width=18cm,height=6.5cm]{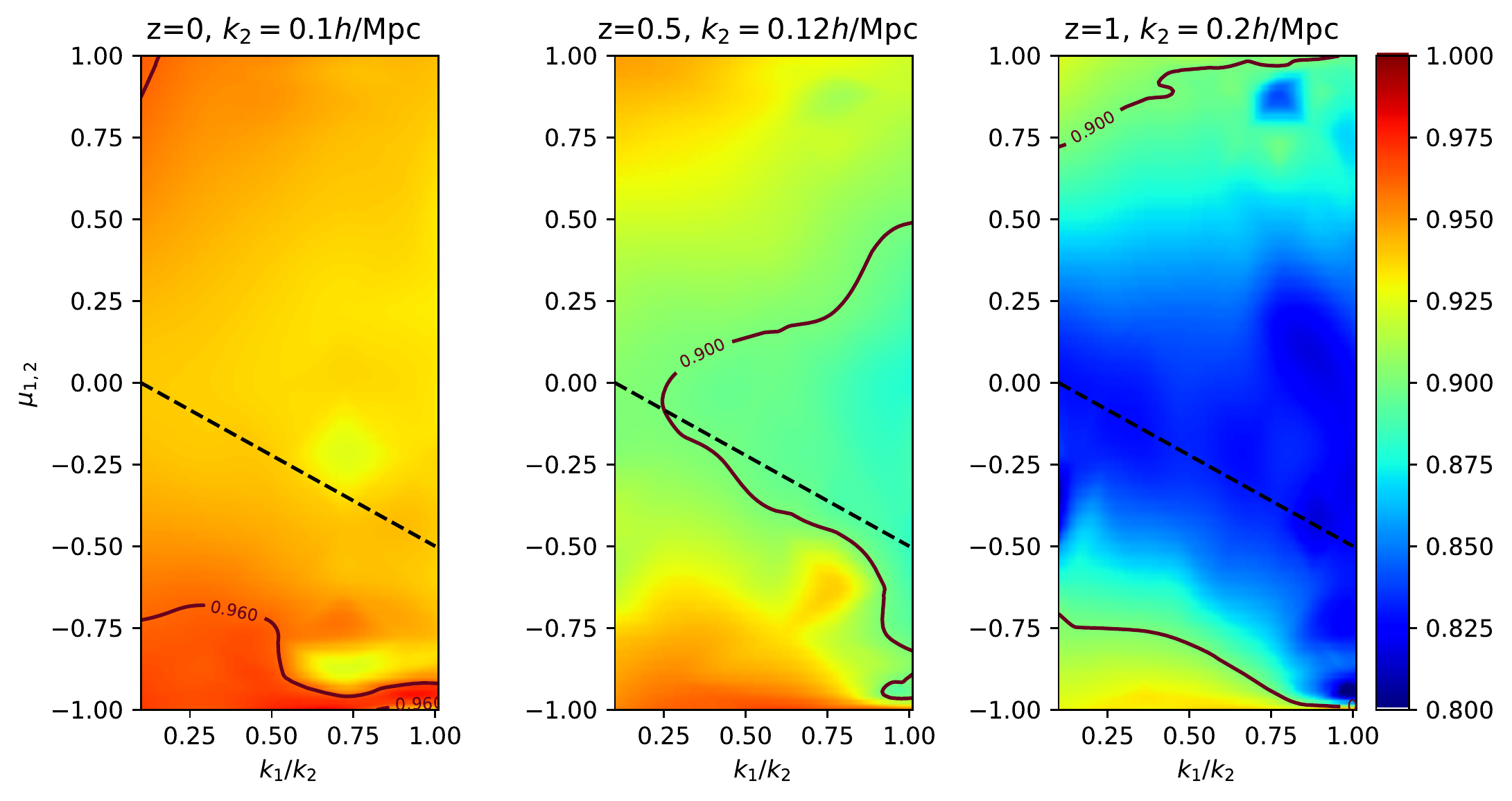}
  \caption[CONVERGENCE ]{Same as the bottom of Fig.~\ref{cont3} ($|\bar{f}_{R0}| = 2.5 \times 10^{-6}$) but at one-loop order.}
 \label{cont4}
\end{figure}

In this section, specifically focusing on the nDGP and $f(R)$ gravity models, we discuss the effects of modifications to gravity, namely those coming from screening and/or which are not directly degenerate with linear power spectrum normalisation, such as $\sigma_8$. In particular, we look at the impact of non-linear MG effects on the shape of the bispectrum and how they vary with redshift. As we are here concerned with the rough magnitude of screening effects and overall triangle shape and redshift dependency we have loosened the accuracy demands of the loop integrations and differential equation solver. This has resulted in some spurious patches noticeable in the one-loop contours of Fig.\ref{cont2} and Fig.\ref{cont4}. These do not change our conclusions or results.

\begin{figure}[H]
\centering
  \includegraphics[width=10cm,height=5.9cm]{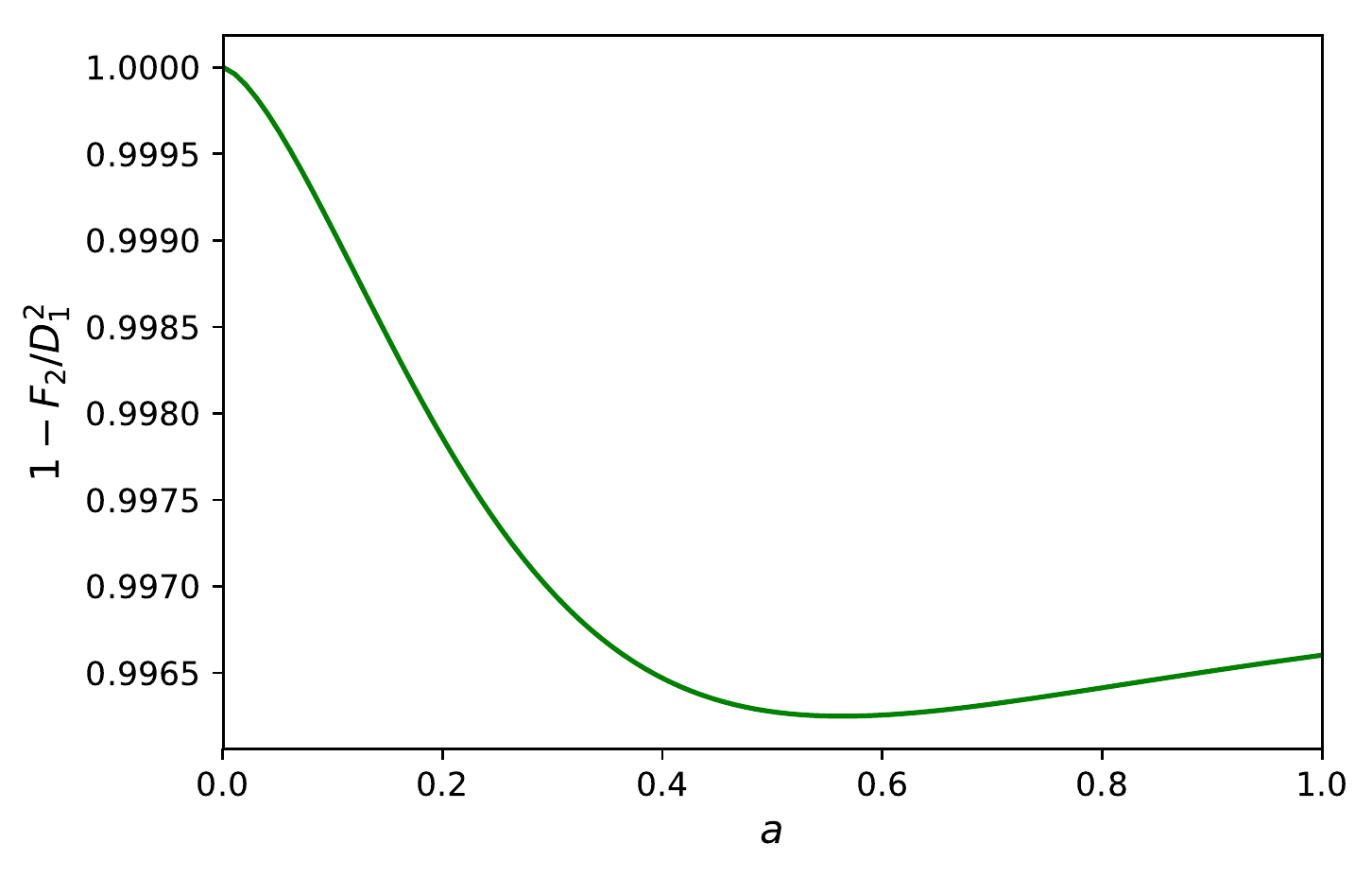}
  \caption[CONVERGENCE ]{Ratio of second-order growth in nDGP to that under the UsA as a function of scale factor $a$.}
\label{growth}
\end{figure}

\begin{figure}[H]
\centering
  \includegraphics[width=18cm,height=6.25cm]{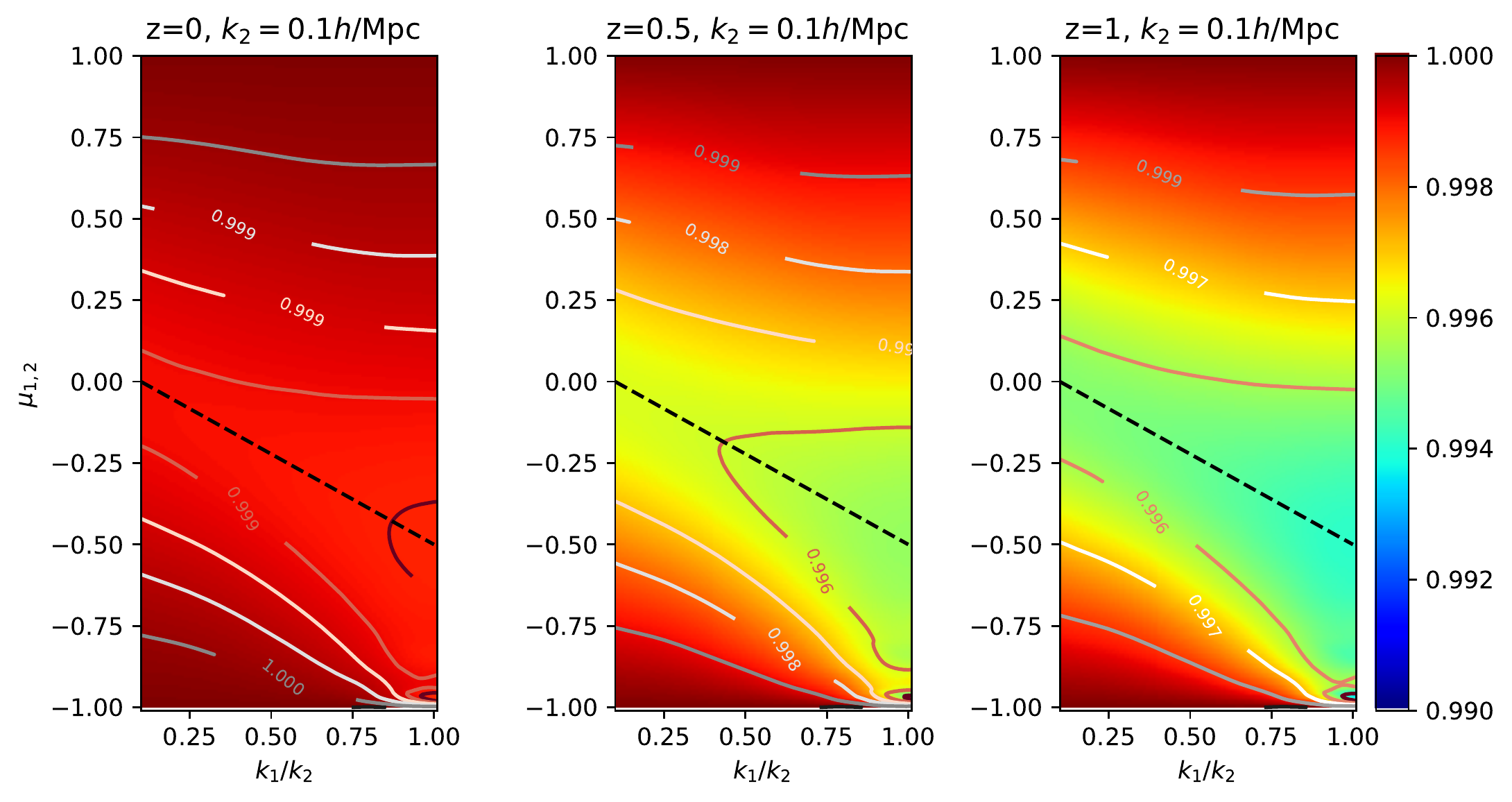}
  \caption[CONVERGENCE ]{ Ratio of tree-level PT prediction of the bispectrum in nDGP to that in nDGP under the UsA, plotted as a function of $k_1/k_2$ and $\mu_{1,2} = (\hat{\bfk}_1 \cdot \hat{\bfk}_2)$, fixing $k_2$ to $0.1\,h\,$$\mbox{Mpc}^{-1}$. The results are shown at $z=0$ (left), $0.5$ (middle), and $1$ (right). The black dashed lines mark the equilateral shape.}
\label{cont1}
\end{figure}

\begin{figure}[H]
\centering
\includegraphics[width=18cm ,height=6.25cm]{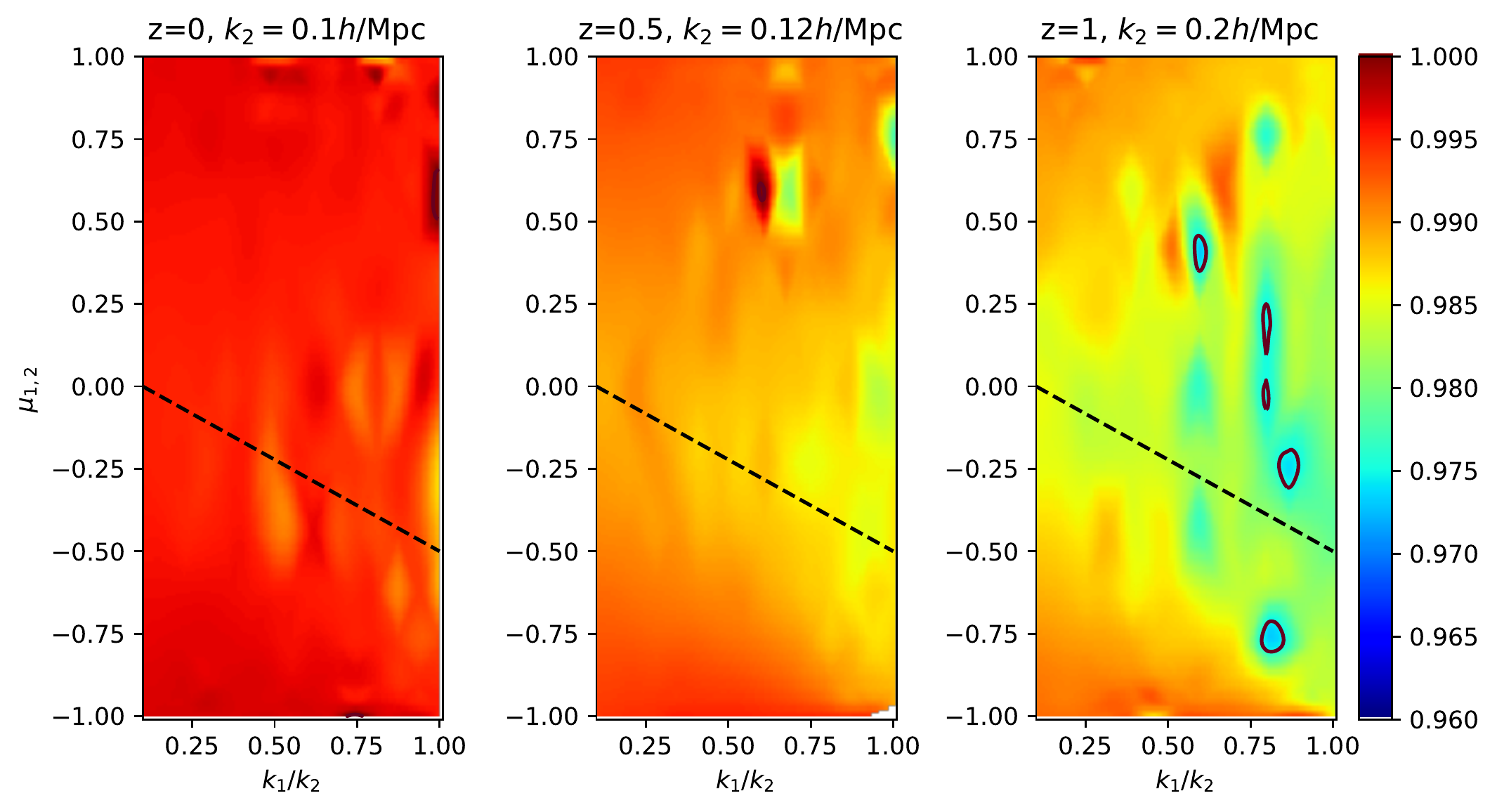}
\caption[CONVERGENCE ]{Same as Fig.\ref{cont1} but results shown are at one-loop order. Here, the wavenumber $k_2$ in each panel is chosen to be $0.1\,h$$\mbox{Mpc}^{-1}$ (left), $0.12\,h$$\mbox{Mpc}^{-1}$ (middle), and $0.2\,h$$\mbox{Mpc}^{-1}$ (right).}
\label{cont2}
\end{figure}

Consider first the bispectrum in nDGP as a prototypical example of MG with Vainshtein screening. At tree-level order, it is shown in Refs.~\cite{Takushima:2013foa,Hirano:2018uar} that Horndeski theories involving Vainshtein screening generally predict an angular-dependent modification through the second-order PT kernel [see Eq.~(\ref{horndeskif2}) with $\kappa=1$], with scale dependence arising from $\bfk_3 = -\bfk_1 - \bfk_2$ terms in the permutations of Eq.~(\ref{tree2}). The second-order kernel in the nDGP model is given by \cite{Koyama:2009me}
\begin{equation}
F_2(\bfk_1,\bfk_2;a) = F_1(a)^2 \Big[ F_2^{\rm GR}(\bfk_1,\bfk_2)+ \frac{F_{2}(a)}{F_1(a)^2}(1-\mu_{1,2}^2) \Big],
\end{equation}
with $F_2^{\rm GR}(\bfk_1,\bfk_2)$ being the second-order PT kernel in GR in the EdS approximation. $F_1$ and $F_2$ are the linear and second-order growth functions in nDGP, and $\mu_{1,2} \equiv \hat{\bfk}_1 \cdot \hat{\bfk}_2$. In Fig.~\ref{growth}, we plot the time evolution of the ratio $F_2(a)/F_1(a)^2$. This highlights the features seen in Fig.~\ref{cont1} where we show the ratio of the tree-level bispectrum in nDGP to that with the UsA (i.e. $F_{2}(a)=0$) given at $z=0$ (left), $0.5$ (middle) and $1$ (right). The results are then plotted as a function of $\mu_{1,2}$ and $k_1/k_2$, fixing $k_2$ to $0.1\,h$$\mbox{Mpc}^{-1}$. Fig.~\ref{growth} shows a purely non-linear modification of gravity valid at tree-level order. This gives a rough idea of the significance of screening effects on the bispectrum and the optimal triangular shape to probe gravity. As deduced from Fig.~\ref{growth}, the modification to gravity in nDGP becomes larger at higher redshift, and is maximal at $\mu_{1,2} \approx -0.5$ and $k_1 \approx k_2$,  corresponding to the equilateral shape, marked by  black dashed lines in Fig.\ref{cont1}. 

Fig.~\ref{cont1} illustrates a generic feature of the bispectrum shape in Horndeski theories with Vainshtein screening. The screening signal in nDGP is quite small, just $\sim 0.5\%$ even at $z=1$, but the redshift dependence and magnitude of the signal are model-dependent, characterized by the parameter $\lambda(a)$ in Eq.~(\ref{horndeskif2}). Further, beyond tree-level order, non-linear modification is highly model-dependent, and characteristic features in the bispectrum shape cannot be simply characterised by a single parameter. Nevertheless, in the presence of screening, one naively expects that the characteristic shape dependence seen at tree-level will tend to be erased at one-loop order. 

Fig.~\ref{cont2} shows the same ratio as in Fig.~\ref{cont1} but at one-loop order using the numerical PT approach. Unlike the tree-level predictions, the shape dependence of the bispectrum varies with scale. Hence, as increasing the redshift from $z=0$ to $1$, we choose $k_2 = 0.1$, $0.12$, and $0.2\,h$$\mbox{Mpc}^{-1}$ (from left to right). As anticipated, the shape dependence seen in the tree-level prediction is mostly washed out. Indeed we see a clearer scale-dependence of the signal, and the magnitude of the screening signal reaches up to $\sim 2.5\%$. 

Consider next $f(R)$ gravity. Fig.~\ref{cont3} plots the ratio of the tree-level bispectrum in $f(R)$ to that in GR. Here, we particularly show the cases with model parameter $|\bar{f}_{R0}| = 1\times 10^{-4}$ (top) and $|\bar{f}_{R0}| = 2.5\times10^{-6}$ (bottom). In contrast to the nDGP model, $f(R)$ gravity involves the chameleon-type screening, with which the scale-dependent enhancement of the linear growth is realised. This is also manifest in most shapes in the tree-level bispectrum, where the ratio to GR becomes significantly larger than unity. While the results in the $|\bar{f}_{R0}| = 1\times 10^{-4}$ case exhibit an extremely large deviation ($\gtrsim30$\%) even within the validity range of SPT, the reasonably small value $|\bar{f}_{R0}| = 2.5\times10^{-6}$, consistent with observations (e.g. \cite{Smith:2009fn,Xu:2014wda}, see also \cite{Jain:2012tn} for a tighter constraint), the enhancement of the ratio becomes rather mild. Still, we see the same trend, and the deviation from GR is maximal at $\mu_{1,2} = 1$, corresponding to $k_3 = k_1 + k_2$. 

However, including the one-loop contributions drastically changes the structure of the shape dependence, shown in Fig.~\ref{cont4}, where we present only the results with $|\bar{f}_{R0}|= 2.5\times10^{-6}$. Again, as in Fig.\ref{cont2}, we set the wavenumber $k_2$ to $0.1$, $0.12$, and $0.2\,h$$\mbox{Mpc}^{-1}$ from left to right panels. The resultant shape dependence looks similar to that in nDGP at one-loop order, with the equilateral shape again giving a maximal deviation. One notable point may be that the amplitude of the bispectrum is now suppressed in comparison to GR, in contrast to the enhancement seen at tree level (see Fig.\ref{cont3}). This would be ascribed to the effect of the screening mechanism, but the magnitude of the suppression seems a bit larger than expected, since the non-linear screening is supposed to be not too effective in the weakly non-linear regime. There might also be the possibility of a break down of SPT even at large scales, however, the qualitative features seen in Fig.~\ref{cont4} would remain the same. At least, one can say that the trend seen in the tree-level prediction generally disappears, and the structure of the shape dependence tends to be the same, although there still remains a non-negligible amount of deviation, which could be a clue to a promising probe of gravity using the bispectrum. 


\section{Conclusion} 
In this paper we have presented an extension of Ref.~\cite{Bose:2016qun} to three-point statistics, specifically a tool to numerically calculate the standard perturbation theory (PT) prediction for the one-loop matter bispectrum. We considered four representative models, namely $\Lambda$CDM, nDGP, $f(R)$ and the phenomenological dark scattering momentum exchange model. In the latter case we consider the phantom model with equation of state parameter of dark energy $w=-1.1$. We have validated the code for standard PT (SPT) calculations against a set of N-body simulations. In the $\Lambda$CDM and nDGP cases, these numerical PT results are also compared with analytic PT predictions involving approximations and/or simplifications as well as  fitting formulas.  Our results are consistent with those previously obtained in $\Lambda$CDM for one-loop bispectra (e.g. \cite{Hashimoto:2017klo}) and for one-loop power spectra (e.g. \cite{Bose:2018orj,Fasiello:2016qpn}).
\newline
\newline
Our important findings for one-loop bispectra are summarized as follows:

\begin{itemize}
 \item  Including one-loop contributions offers a large gain in accuracy over the leading-order (tree-level) predictions in all models considered. The accuracy of one-loop bispectra is comparable to the fitting formulas at higher redshift ($z\gtrsim0.5$) in the quasi linear regime ($k \lesssim 0.15$ - $0.25\,h$\,Mpc$^{-1}$) and the one-loop SPT bispectrum prediction reproduces well the simulations at a relatively wider range than that of the power spectrum.  

\item Analytic PT treatment involving approximations/simplifications generally  
produces a percent level deviation from the numerical PT approach. While the Einstein-de Sitter approximation just gives a sub-percent error and hence can be safely applied, the omission of screening effects at higher-order can produce an error that reaches the percent level, which may be of concern to upcoming surveys, although the actual impact would depend on survey errors and other nuisance parameters. 

\item Characteristic shape dependence seen in modified gravity models, which appears at tree-level order, tends to be erased as we move to lower redshift in the one-loop SPT prediction. For instance, in nDGP, taken as a representative model of the Horndeski class, the tree-level bispectrum exhibits a clear maximal deviation from GR in the equilateral configuration. This is qualitatively the same in $f(R)$ gravity. At one-loop order, however, the shape dependence drastically changes, and becomes similar in both nDGP and $f(R)$ gravity, although the magnitude of the deviation depends on the specific model. Interestingly, the equilateral shape still shows the maximal deviation, and its magnitude is up to $4$ times as large as the signal exhibited in the tree-level prediction, indicating that one-loop bispectrum could be a promising probe of modified gravity. 
\end{itemize}

The numerical PT framework presented here naturally finds many extensions available to PT. For example, one can include prescriptions that improve the poor-convergence properties in SPT calculation. One example would be the inclusion of resummation such as multi-point propagator expansion \cite{Bernardeau:2008fa,Bernardeau:2011dp,Taruya:2012ut}. Also, the effective field theory of large scale structure  \cite{Baumann:2010tm,Carrasco:2012cv} has been extended to the bispectrum \cite{Angulo:2014tfa,Baldauf:2014qfa}, which could be useful in extracting valuable information from small scales. In confronting observations, CMB lensing can offer a relatively clean probe of gravity \cite{Namikawa:2018erh}, for which application of our pipeline is rather straightforward. As a first step, in a future work, we shall examine simulated lensing data to further investigate some of the claims proposed here. Further, the redshift-space bispectrum has recently been measured in the BOSS survey \cite{Pearson:2017wtw,Sugiyama:2018yzo} and a promising redshift-space bispectrum model has also been proposed at one-loop order in \cite{Hashimoto:2017klo}. Extending our treatment to redshift space is thus another interesting avenue. However, this would involve some severe numerical optimisations to the code used in this paper, since an additional two-dimensional integral would need to be performed to obtain the bispectrum multipoles (e.g., \cite{Sugiyama:2018yzo}). Further, substantial optimisations are also needed in order to apply our numerical one-loop bispectrum to the parameter estimation analysis, typically using the Markov Chain Monte Carlo technique.  One may also consider gravitational and dark energy effects on the 3 point correlation function (see \cite{Slepian:2016weg} for a recent model for GR). Recently progress has been made in methods to estimate and measure this in redshift space \cite{Slepian:2017lpm,Friesen:2017acf}, making it another interesting statistic relevant for upcoming surveys, especially as it provides a means of overcoming systematics typical of the bispectrum. These points are currently within the authors' focus.

On top of this we have the issue of tracer bias. Recently a fully comprehensive bias model for the one-loop bispectrum has also been derived based on the bias expansion approach \cite{Desjacques:2018pfv}. This primes an investigation into the constraining power of the one-loop redshift space galaxy bispectrum for non-standard models of cosmology, and if moving beyond consistency tests of $\Lambda$CDM can be achieved with future spectroscopic surveys. On this note, there is still the major issue of the covariance between redshift-space multipoles which has been mostly studied for the Gaussian case  \cite{Scoccimarro:1997st,Scoccimarro:2003wn} and has been restricted to GR  \cite{Takada:2003ef,Kayo:2012nm,Sato:2013mq,Chan:2016ehg}. We leave the study of this in theories beyond $\Lambda$CDM to a future work.
\newpage

\section*{Acknowledgments}
\noindent BB is JSPS International Research Fellow (PE17043) and acknowledges support from JSPS. This work was supported in part by MEXT/JSPS KAKENHI Grant Number JP15H05899 and JP16H03977 (AT). The authors would like to thank Marco Baldi for providing the Dark Scattering simulation data and Hans Winther for providing the $\Lambda$CDM and nDGP simulation data. We would also like to thank Takahiro Nishimichi for providing the packages for fast bispectrum measurements from the simulation snapshots. Kazuya Koyama is also thanked for useful discussions. The $f(R)$ simulations were run on the {\tt Sciama} super computer of the University of Portsmouth. 

\appendix
\section{Numerical Accuracy and Timing Results}
\label{appendix:numerical_accuracy}
In this appendix we give some details on numerical accuracy of the approach described in Sec. II C. The differential equations Eq.(\ref{evoleqn:fn}) and Eq.(\ref{evoleqn:gn}) must be solved at the bottom level of the 3 dimensional integral of the one-loop integrals [for example in Eq.(\ref{fourtho})]. For the integration, we employ an adaptive 15-point Gauss-Kronrod rule. The accuracy of the integral and the differential equation solver are tuned in Sec. III so that the numerical result agrees to the percent level with the Einstein-de Sitter (cosmology with $\Omega_m = 1$) analytic result, which is exact in this case. Small increases in the relative error of the integral and differential equation solver lead to significant time costs, making this choice very dependent on the accuracy required for the given analysis. We note that in Sec. IV we loosen the accuracy demands as we look to demonstrate trends and rough magnitudes. For the bispectrum computations we have employed a rough adaptive method which demands low accuracy in the integrator at large scales and high accuracy at small scales. This has given the best overall compromise between accuracy and time cost. Refinement of this method will be necessary for future analyses. 

\begin{figure}[H]
\centering
  \includegraphics[width=15cm,height=5cm]{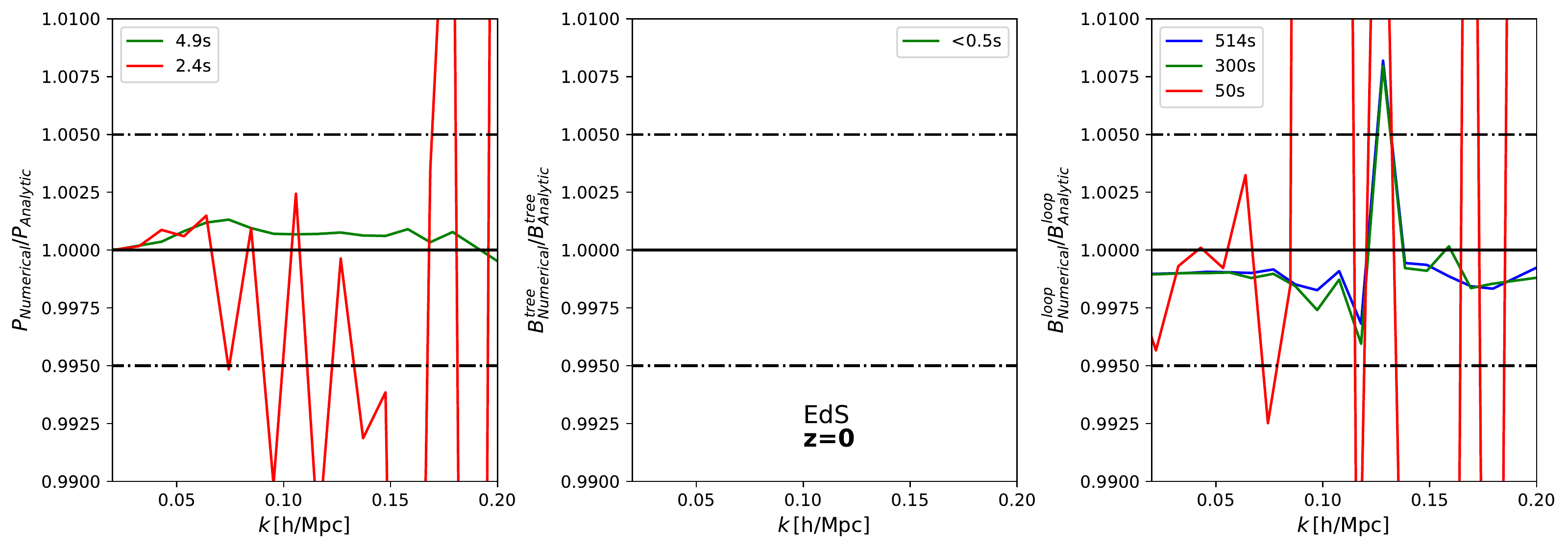}
  \caption[CONVERGENCE ]{Ratio of numerically computed one-loop power spectrum (left), tree level equilateral bispectrum (middle) and one-loop equilateral bispectrum (right) to the EdS analytic spectra for varying levels of integrator relative error. Time quoted in legend refers to the total time taken to compute all 20 $k$ values used in the plots. All timing results were obtained on a MacBook Pro laptop computer, with a 2.52 GHz Intel Core 2 Duo processor and running on Mac OS X version 10.6.8.}
\label{checker}
\end{figure}

In Fig.\ref{checker} we show some timing results with varying levels of relative error in the 3D-integration at $z=0$. Specifically, we show the ratio between the EdS numerical PT computation and EdS analytic solution for the one-loop power spectrum (left), tree level equilateral shape bispectrum (middle) and one-loop equilateral shape bispectrum (right).  The differential equation solver's accuracy is tuned so that the tree level bispectrum result is sub $0.1\%$. The different curves then show different levels of fixed accuracy in the loop integral. The green curve shows the level adopted in this paper, with the red adopting a lower accuracy and the blue a higher one (right plot only). The green curve's numerical inaccuracies shown in the one-loop power spectrum are sub $0.1\%$ but are noticeably larger in the one-loop bispectrum (right panel), albeit still sub $1\%$. These are likely to come from not properly treating the $|\bfk-\bfq|$ type divergences in the loop integrals, where $\bfq$ is the integrated wave vector. Implementing the fully IR-safe integral \cite{Baldauf:2014qfa} is numerically challenging as this comes at a significant time cost to the computation. This is a current focus of the authors. For the accuracy demanded in this paper, and further for ongoing and future surveys aiming at using the bispectrum, the errors induced by our `IR-unsafe' approach may be acceptable. We note that the tree level accuracy is such that the deviation is sub $0.01\%$ and so the green curve lies directly under the black line denoting a ratio of unity.

\section{Separable Solutions in nDGP: Fourth Order Kernels} 
\label{appendix:kernel_F4_in_DGP}

Here we present the 4th order kernel in nDGP gravity under the EdS approximation. Using Eq.(\ref{eq:Perturb1}) and Eq.(\ref{eq:Perturb2}) we can write the following 2nd order differential equation for the density contrast at 4th order
\begin{align}
&\ddot{\delta} + 2H\dot{\delta} - \frac{\kappa \rho_m}{2}\left(1+\frac{1}{3\beta(a)}\right) = \int \frac{d^3k_1d^3k_2d^3k_3d^3k_4}{(2\pi)^{12}} \delta_D(\bfk-\bfk_{1234})  \nonumber \\ 
& \times \left[  \frac{1}{2}\beta(\bfk_1,\bfk_{234}) \theta_1(\bfk_1) \theta_3(\bfk_2,\bfk_3,\bfk_4) + \frac{1}{2}\beta(\bfk_{123},\bfk_{4}) \theta_3(\bfk_1,\bfk_2,\bfk_3) \theta_1(\bfk_4) +\frac{1}{2}\beta(\bfk_{12},\bfk_{34}) \theta_2(\bfk_1,\bfk_2) \theta_2(\bfk_3,\bfk_4) \right.   \nonumber \\ 
&  -(2H^2 + \dot{H}) \Big[ \alpha(\bfk_1,\bfk_{234}) \theta_1(\bfk_1) \delta_3(\bfk_2,\bfk_3,\bfk_4)  + \alpha(\bfk_{123},\bfk_{4}) \theta_3(\bfk_1,\bfk_2,\bfk_3)  \delta_1(\bfk_4)  \nonumber \\  & \qquad  \qquad \qquad + \alpha(\bfk_{12},\bfk_{34}) \theta_2(\bfk_1,\bfk_2) \delta_2(\bfk_3,\bfk_4) \Big]  \nonumber \\ 
& -H\left[\alpha(\bfk_1,\bfk_{234}) \dot{\theta}_1(\bfk_1) \delta_3(\bfk_2,\bfk_3,\bfk_4) + \alpha(\bfk_{123},\bfk_{4}) \dot{\theta}_3(\bfk_1,\bfk_2,\bfk_3)  \delta_1(\bfk_4) \right.   \nonumber \\ 
&\qquad \quad   +\alpha(\bfk_1,\bfk_{234}) {\theta}_1(\bfk_1) \dot{\delta}_3(\bfk_2,\bfk_3,\bfk_4) + \alpha(\bfk_{123},\bfk_{4}) {\theta}_3(\bfk_1,\bfk_2,\bfk_3)  \dot{\delta}_1(\bfk_4) \nonumber \\ 
&\qquad  \quad \left.  + \alpha(\bfk_{12},\bfk_{34}) \dot{\theta}_2(\bfk_1,\bfk_2) \delta_2(\bfk_3,\bfk_4)  +  \alpha(\bfk_{12},\bfk_{34}) {\theta}_2(\bfk_1,\bfk_2) \dot{\delta}_2(\bfk_3,\bfk_4)\right]   \nonumber \\
& + H^2 \left[ \gamma_2(\bfk_1,\bfk_{234}) \delta_1(\bfk_1) \delta_3(\bfk_2,\bfk_3,\bfk_4) + \gamma_2(\bfk_{123},\bfk_{4}) \delta_3(\bfk_1,\bfk_2,\bfk_3) \delta_1(\bfk_4) +   \gamma_2(\bfk_{12},\bfk_{34}) \delta_2(\bfk_1,\bfk_2) \delta_2(\bfk_3,\bfk_4) \right. \nonumber \\ 
&  + \gamma_3(\bfk_1,\bfk_2,\bfk_{34}) \delta_1(\bfk_1) \delta_1(\bfk_2) \delta_2(\bfk_3,\bfk_4) + \gamma_3(\bfk_1,\bfk_{23},\bfk_{4}) \delta_1(\bfk_1) \delta_2(\bfk_2,\bfk_3) \delta_1(\bfk_4) \nonumber \\ & \left.+  \gamma_3(\bfk_{12},\bfk_3,\bfk_{4}) \delta_2(\bfk_1,\bfk_2) \delta_1(\bfk_3) \delta_1(\bfk_4)   + \gamma_4(\bfk_1,\bfk_2,\bfk_3,\bfk_4) \delta_1(\bfk_1) \delta_1(\bfk_2) \delta_1(\bfk_3) \delta_1(\bfk_4) \right] \Big],
\end{align}
where $\gamma_i$ and $\beta(a)$ are given in Sec.II B.  Assuming the GR solution is given by the EdS approximation we can solve for the scale dependencies and evolution of the DGP part. First we can expand the DGP density contrast into the EdS solution and the part arising from the extra vertices in DGP: $\delta_{2-4} = \delta_{{\rm EdS},2-4} + \delta_{dgp,2-4}$ and similarly for $\theta_{2-4}$. The expressions for the 2nd and 3rd order density contrasts can be found in Appendix B of \cite{Koyama:2009me}. We are left with the following form of the fourth order density contrast in nDGP 
\begin{align}
& F_{\rm 4,DGP}(\bfk_1,\bfk_2,\bfk_3,\bfk_4;a) =   F_{\rm 4,GR}^{\rm EdS}((\bfk_1,\bfk_2,\bfk_3,\bfk_4;a) \nonumber \\ &
 + \frac{1}{4} \Big[  {\bf Z}(\bfk_2,\bfk_3,\bfk_4) \cdot \Big( {\bf A}_4(a)  \beta(\bfk_1,\bfk_{234}) + {\bf D}_4(a)\alpha(\bfk_{1},\bfk_{234})  + {\bf E}_4(a)  \alpha(\bfk_{234},\bfk_{1})  + 2{\bf I}_4(a) (1-\mu_{1,234}^2) \Big)  \nonumber \\ 
& \qquad  + 2 L_4(a) F_3(\bfk_2,\bfk_3,\bfk_4) (1-\mu_{1,234}^2)  + \mbox{3 perms} \Big] \nonumber \\ 
& + \frac{1}{6} \Big[ \frac{1}{2} B_4(a) (1-\mu_{1,2}^2) G_2(\bfk_3,\bfk_4)  \beta(\bfk_{12},\bfk_{34}) +  \frac{1}{2} C_4(a) (1-\mu_{1,2}^2) (1-\mu_{3,4}^2) \beta(\bfk_{12},\bfk_{34})  \nonumber \\ 
& \qquad+ F_4(a)  G_2(\bfk_1,\bfk_2)(1-\mu_{3,4}^2) \alpha(\bfk_{12},\bfk_{34})  + G_4(a)  (1-\mu_{1,2}^2) F_2(\bfk_3,\bfk_4) \alpha(\bfk_{12},\bfk_{34})  \nonumber \\ 
& \qquad+ H_4(a)  (1-\mu_{1,2}^2) (1-\mu_{3,4}^2)  \alpha(\bfk_{12},\bfk_{34})  + J_4(a)  F_2(\bfk_1,\bfk_2) (1-\mu_{3,4}^2) (1-\mu_{12,34}^2)   \nonumber \\  & \qquad+ K_4(a)  (1-\mu_{1,2}^2)(1-\mu_{3,4}^2) (1-\mu_{12,34}^2)   + L_4(a)  F_2(\bfk_1,\bfk_2) F_2(\bfk_3,\bfk_4) (1-\mu_{12,34}^2)  + \mbox{ 5 perms} \Big]   \nonumber \\ 
& + \frac{1}{12} \Big[ N_4(a)   (1-\mu_{3,4}^2)  (1-\mu_{1,234}^2) (1-\mu_{2,34}^2)  + O_4(a)   F_2(\bfk_3,\bfk_4)  (1-\mu_{1,234}^2) (1-\mu_{2,34}^2)  + \mbox{11 perms} \Big], \label{f4dgp} 
\end{align}
\begin{align}
& G_{\rm 4,DGP}(\bfk_1,\bfk_2,\bfk_3,\bfk_4;a) =   G_{\rm 4,GR}^{\rm EdS}((\bfk_1,\bfk_2,\bfk_3,\bfk_4;a) \nonumber \\ &
 - \frac{1}{4} \Big[  {\bf Z}(\bfk_2,\bfk_3,\bfk_4) \cdot \Big( \dot{{\bf A}}_4(a)  \beta(\bfk_1,\bfk_{234}) + \Big[\dot{{\bf D}}_4(a) - \frac{\dot{D}(a)}{H(a)} {\bf D}_\delta(a) \Big] \alpha(\bfk_{1},\bfk_{234})  + \Big[\dot{{\bf E}}_4(a) - F_1(a)  {\bf D}_\theta(a)  \Big] \alpha(\bfk_{234},\bfk_{1})  \nonumber \\ & \qquad + 2\dot{{\bf I}}_4(a) (1-\mu_{1,234}^2) \Big)  
  + 2 \dot{L}_4(a) F_3(\bfk_2,\bfk_3,\bfk_4) (1-\mu_{1,234}^2)  + \mbox{3 perms} \Big] \nonumber \\ 
& - \frac{1}{6} \Big[ \frac{1}{2} \dot{B}_4(a) (1-\mu_{1,2}^2) G_2(\bfk_3,\bfk_4)  \beta(\bfk_{12},\bfk_{34}) +  \frac{1}{2} \dot{C}_4(a) (1-\mu_{1,2}^2) (1-\mu_{3,4}^2) \beta(\bfk_{12},\bfk_{34})  \nonumber \\ 
& \qquad+ \Big[\dot{F}_4(a) - \frac{F_2(a)\dot{D}_1(a)F_1(a)}{H(a)} \Big] G_2(\bfk_1,\bfk_2)(1-\mu_{3,4}^2) \alpha(\bfk_{12},\bfk_{34})  + \Big[\dot{G}_4(a)-\frac{\dot{F}_2(a)F_1(a)^2}{H(a)}\Big]  (1-\mu_{1,2}^2) F_2(\bfk_3,\bfk_4) \alpha(\bfk_{12},\bfk_{34})  \nonumber \\ 
& \qquad+ \Big[\dot{H}_4(a)-\frac{\dot{F}_2(a)F_2(a)}{H(a)} \Big]  (1-\mu_{1,2}^2) (1-\mu_{3,4}^2)  \alpha(\bfk_{12},\bfk_{34})  + \dot{J}_4(a)  F_2(\bfk_1,\bfk_2) (1-\mu_{3,4}^2) (1-\mu_{12,34}^2)   \nonumber \\  & \qquad+ \dot{K}_4(a)  (1-\mu_{1,2}^2)(1-\mu_{3,4}^2) (1-\mu_{12,34}^2)   + \dot{L}_4(a)  F_2(\bfk_1,\bfk_2) F_2(\bfk_3,\bfk_4) (1-\mu_{12,34}^2)  + \mbox{ 5 perms} \Big]   \nonumber \\ 
& - \frac{1}{12} \Big[ \dot{N}_4(a)   (1-\mu_{3,4}^2)  (1-\mu_{1,234}^2) (1-\mu_{2,34}^2)  + \dot{O}_4(a)   F_2(\bfk_3,\bfk_4)  (1-\mu_{1,234}^2) (1-\mu_{2,34}^2)  + \mbox{11 perms} \Big],
\end{align}
 where we have also presented the velocity divergence kernel for completion \footnote{To obtain the fourth order velocity divergence kernel one can simply use Eq.(\ref{eq:Perturb1}).}. All kernels in the above expressions are symmetrised and 
\begin{align}
{\bf Z}(\bfk_1,\bfk_2,\bfk_3)  \equiv \Big[ &C_{sym}(\bfk_1,\bfk_2,\bfk_3),F_{sym}(\bfk_1,\bfk_2,\bfk_3),I_{sym}(\bfk_1,\bfk_2,\bfk_3),J_{sym}(\bfk_1,\bfk_2,\bfk_3), \nonumber \\ & K_{sym}(\bfk_1,\bfk_2,\bfk_3),L_{sym}(\bfk_1,\bfk_2,\bfk_3)  \Big],
\end{align}
where the vector's components are the additional 3rd order kernels in DGP \cite{Koyama:2009me}. The evolution factors are then given by 
\begin{align}
\hat{\mathcal{L}} {\bf A}_4 &= \dot{F_1} {\bf D}_\theta, \\
\hat{\mathcal{L}} {\bf D}_4 &= \frac{\kappa \rho_m}{2} \left(1+\frac{1}{3\beta(a)} \right) F_1 {\bf D}_\delta + \dot{D}_1 \dot{{\bf D}}_\delta, \\
\hat{\mathcal{L}} {\bf E}_4 &= \frac{\kappa \rho_m}{2} \left(1+\frac{1}{3\beta(a)} \right) F_1{\bf D}_\delta + \dot{D}_1 \dot{{\bf D}}_\delta - {\bf Y}, \\
\hat{\mathcal{L}} {\bf I}_4 &= - \frac{H_0^2}{24 \beta(a)^3 \Omega_{rc}} \left(\frac{\Omega_{m,0}}{a^3}\right)^2 F_1  {\bf D}_\delta , \\
\hat{\mathcal{L}} L_4 &=  - \frac{H_0^2}{24 \beta(a)^3 \Omega_{rc}} \left(\frac{\Omega_{m,0}}{a^3}\right)^2 F_1^4, \\
\hat{\mathcal{L}} B_4 &=  \dot{D}_1 F_1 \dot{F}_2, \\ 
\hat{\mathcal{L}} C_4 &=  \dot{F}_2^2, 
\end{align}
\begin{align}
\hat{\mathcal{L}} F_4 &=  \frac{\kappa \rho_m}{2} \left(1+\frac{1}{3\beta(a)} \right) F_1^2 F_2 + \dot{D}_1^2 F_2 + \dot{D}_1 \dot{F}_2 F_1, \\ 
\hat{\mathcal{L}} G_4 &=  \frac{\kappa \rho_m}{2} \left(1+\frac{1}{3\beta(a)} \right) F_1^2 F_2   - \frac{H_0^2}{24 \beta(a)^3 \Omega_{rc}} \left(\frac{\Omega_{m,0}}{a^3}\right)^2 F_1^4 + 2\dot{D}_1 \dot{F}_2 F_1,  \\ 
\hat{\mathcal{L}} H_4 &=  \frac{\kappa \rho_m}{2} \left(1+\frac{1}{3\beta(a)} \right) F_2^2   - \frac{H_0^2}{24 \beta(a)^3 \Omega_{rc}} \left(\frac{\Omega_{m,0}}{a^3}\right)^2 F_2 F_1^2 +  \dot{F}_2^2 , \\
\hat{\mathcal{L}} J_4 &=  - \frac{2H_0^2}{24 \beta(a)^3 \Omega_{rc}} \left(\frac{\Omega_{m,0}}{a^3}\right)^2 F_2 F_1^2 + \frac{H_0^2}{144 \beta(a)^5 \Omega_{rc}^2} \left(\frac{\Omega_{m,0}}{a^3}\right)^3 F_1^4 , \\
\hat{\mathcal{L}} K_4 &=  - \frac{H_0^2}{12 \beta(a)^3 \Omega_{rc}} \left(\frac{\Omega_{m,0}}{a^3}\right)^2 F_2 F_2 + \frac{H_0^2}{144 \beta(a)^5 \Omega_{rc}^2} \left(\frac{\Omega_{m,0}}{a^3}\right)^3 F_1^2 F_2 \nonumber \\ 
& \quad  -  \frac{H_0^2}{864 \beta(a)^7 \Omega_{rc}^3} \left(\frac{\Omega_{m,0}}{a^3}\right)^4 F_1^4  , \\
\hat{\mathcal{L}} N_4 &=  \frac{2H_0^2}{144 \beta(a)^5 \Omega_{rc}^2} \left(\frac{\Omega_{m,0}}{a^3}\right)^3 F_1^2 F_2 -  \frac{H_0^2}{3456 \beta(a)^7 \Omega_{rc}^3} \left(\frac{\Omega_{m,0}}{a^3}\right)^4 F_1^4  , \\
\hat{\mathcal{L}} O_4 &=  \frac{2H_0^2}{144 \beta(a)^5 \Omega_{rc}^2} \left(\frac{\Omega_{m,0}}{a^3}\right)^3 F_1^4 , 
\end{align}
where $\Omega_{rc} = 1/(4H_0^2 r_c^2)$, $r_c$ being the cross-over scale, and 
\begin{equation}
\mathcal{\hat{L}} \equiv  a^2 H^2 \frac{d^2}{da^2} + aH^2 \left( 3+ \frac{a H'}{H} \right)\frac{d}{da} - \frac{\kappa \rho_m}{2} \left( 1+ \frac{1}{3\beta} \right) \label{lder}, 
\end{equation}
and 
\begin{align}
{\bf D}_\delta &\equiv \Big[ C_3, F_3, I_3,J_3,K_3,L_3 \Big], \\
{\bf D}_\theta &\equiv \Big[ \dot{C}_3, (\dot{F}_3 - \dot{D}_1 F_2) , (\dot{I}_3-F_1\dot{F}_2) ,\dot{J}_3, \dot{K}_3, \dot{L}_3 \Big], \\ 
{\bf Y} &\equiv \Big[0, F_1^2 F_2 \frac{\kappa \rho_m}{2} \left(1+\frac{1}{3\beta(a)} \right)  + F_1 \dot{D}_1 \dot{F}_2 + \dot{D}_1^2 F_2 , \nonumber \\ & \qquad  F_1^2 F_2 \frac{\kappa \rho_m}{2} \left(1+\frac{1}{3\beta(a)} \right)  + 2F_1 \dot{D}_1 \dot{F}_2 , 0, 0, 0\Big]. 
\end{align}
The GR 4th order kernels $F_{\rm 4,GR}^{\rm EdS}(\bfk_1,\bfk_2,\bfk_3,\bfk_4;a)$ and $G_{\rm 4,GR}^{\rm EdS}((\bfk_1,\bfk_2,\bfk_3,\bfk_4;a)$ are standard results and can be derived using Eq.(43) and Eq.(44) of \cite{Bernardeau:2001qr} for example.

\section{Fitting Formula for General Scalar Tensor Theories}
\label{appendix:fitting_formula_bispec}

Here we present the fitting formula for the matter bispectrum proposed in \cite{Namikawa:2018erh} for the beyond Horndeski class of theories. We begin by noting the form for the 2nd order kernel in beyond Horndeski theories within the quasi-static and EdS approximations is given by \cite{Hirano:2018uar}
\begin{equation}
F_2(\bfk_1,\bfk_2;a) = F_1(a)^2 \Big[ \frac{\kappa(a)}{2} \left[ \alpha(\bfk_1,\bfk_2) + \alpha(\bfk_2,\bfk_1) \right] - \frac{2 }{7} \lambda(a) (1-\mu_{1,2}^2) \Big],
\label{horndeskif2}
\end{equation}
where again $\mu_{1,2} = \hat{\bfk}_1 \cdot \hat{\bfk}_2$, $F_1(a)$ is the linear growth factor and $\kappa(a)$ and $\lambda(a)$ are 2nd order time-dependent functions that are theory-dependent. For Horndeski theories we set $\kappa(a)=1$ and for GR $\kappa(a) =\lambda(a) =1$. This expression was given a non-linear extension in \cite{Namikawa:2018erh} based on the GR bispectrum fitting formula presented in \cite{Scoccimarro:2000ee}. We quote this below
\begin{align}
F^{\rm fit}_2(\bfk_1,\bfk_2;a) = &  F_1(a)^2 \Big[\left(\kappa(a)- \frac{2}{7}\lambda(a)\right)\bar{a}(k_1,a)\bar{a}_2(k_2,a) \nonumber \\ & \qquad  \qquad + \frac{\kappa(a)}{2} \mu_{1,2} \frac{k_1^2 + k_2^2}{k_1 k_2}\bar{b}(k_1,a)\bar{b}_2(k_2,a)  + \lambda(a) \frac{2}{7} \mu_{1,2}^2\bar{c}(k_1,a)\bar{c}_2(k_2,a) \Big] ,
\label{fittingbh}
\end{align}
where the non-linear prescription is through the following functions  
\begin{align} 
\bar{a}(k,a) &= \frac{1+[\sigma_8(a)]^{a_6} \sqrt{0.7Q(n(k))} (qa_1)^{n(k)+a_2}}{1+(qa_1)^{n(k)+a_2}}, \\
 \bar{b}(k,a) &= \frac{1+0.2a_3 (n(k)+3)(qa_7)^{n(k)+3+a_8} }{1+(qa_7)^{n(k)+3.5+a_8}}, \\
 \bar{c}(k,a) &= \frac{1+[4.5a_4/(1.5+(n(k)+3)^4) ](n(k)+3)(qa_5)^{n(k)+3+a_9} }{1+(qa_5)^{n(k)+3.5+a_9}}, 
\end{align}
with 
\begin{equation}
 Q(x) = (4-2^x)/(1+2^{x+1}), \qquad \mbox{and} \qquad  n(k) = \frac{d\log{P_L(k')}}{d\log{k'}} |_k.
 \end{equation}
 The various other quantities are $q=k/k_{NL}$, where $k_{NL}$ is the scale where non-linearities start to become important, determined by $k^3_{NL} P_L(k_{NL})/(2\pi^2) = 1$, and $a_{1-9}$ are constants that are determined by fitting to N-body simulations. We use the values found in \cite{GilMarin:2011ik} which are determined from GR simulations, thus all screening information in this approach is encoded solely in the modification of the $F_2$ kernel given in Eq.(\ref{horndeskif2}). The prescription for the non-linear bispectrum takes the form 
\begin{equation}
B^{fit}(k_1,k_2,\theta;a) = 2 F^{fit}_2(\bfk_1,\bfk_2;a) P_{NL}(k_1;a) P_{NL}(k_2;a) + 2 \mbox{perms} (\bfk_1 \leftrightarrow \bfk_2 \leftrightarrow \bfk_3) \label{bfit}, 
\end{equation}
where $P_{NL}$ is some prescription for the non-linear matter power spectrum. As in \cite{GilMarin:2011ik}, we employ the {\tt halofit} model prescription for $P_{NL}$ \cite{Smith:2002dz,Takahashi:2012em}, and simply replace the linear growth factors with the modified linear growth factors. Lastly, one must treat spurious oscillations that arise due to the oscillations in $n(k)$ coming from baryon acoustic features. In \cite{GilMarin:2011ik} they employ a somewhat involved method that splines $n(k)$ through the middle of each oscillation. Here we take a simpler route and use a no-wiggle spectrum proposed in Eq.(2.47) of \cite{delaBella:2017qjy}. This approach effectively filters out the baryon acoustic oscillatory features but preserves the amplitude and broadband shape of the spectrum. 
\newline
\newline
For nDGP $\kappa(a)=1$ and $\lambda(a) = (1- \frac{7}{2}\frac{F_{2}(a)}{F_1(a)^2})$, where $F_{2}(a)$ is the 2nd order growth factor in nDGP, given by solving the following evolution equation 
 \begin{equation}
 \mathcal{\hat{L}} F_{2}(a) = -\frac{H_0^2}{24 \Omega_{rc}} \left(\frac{\Omega_{m,0}}{a^3}\right) F_1(a)^2,
 \end{equation}  
where $\mathcal{\hat{L}}$ is given by Eq.(\ref{lder}). For the DS model we set $\kappa=\lambda=1$ and use the linear growth factor found by solving the linear version of Eq.(\ref{evoleqn:fn}). 

\bibliography{mybib}{}
\end{document}